\documentclass[preprint,10pt]{elsarticle}

\usepackage{amsmath,amssymb}

\usepackage{booktabs}
\usepackage{tikz}
\usetikzlibrary{positioning, arrows.meta}
\usepackage{array}
\usepackage{rotating}
\usepackage{siunitx}
\usepackage{tabularx}
\usepackage{graphicx}
\usepackage{subcaption}
\usepackage{hyperref}
\hypersetup{
    colorlinks=true,
    linkcolor=blue,
    citecolor=blue,
    urlcolor=blue
}

\biboptions{sort&compress}

\journal{Ocean engineering}

\begin{document}

\begin{frontmatter}

\title{A Methodology for Characterizing Underwater Radiated Noise from Submerged Electric Vehicles in a Coastal Environment: An AUV Test Case}

\author[1]{Mark Shipton}

\author[3]{Amir Boag}

\author[1,2]{Roee Diamant\corref{cor1}}
\ead{roee.d@univ.haifa.ac.il}

\cortext[cor1]{Corresponding author}

\affiliation[1]{
  organization={Department of Marine Technologies, University of Haifa},
  city={Haifa},
  country={Israel}
}

\affiliation[2]{
  organization={Faculty of Electrical Engineering and Computing, University of Zagreb},
  city={Zagreb},
  country={Croatia}
}

\affiliation[3]{
  organization={School of Electrical and Computer Engineering, Tel Aviv University},
  city={Tel Aviv},
  country={Israel}
}

\begin{abstract}
\footnotesize
Submerged electric vehicles (SEVs), including autonomous underwater vehicles (AUVs), remotely operated vehicles, and diver propulsion systems, may radiate distinct tonal, harmonic, and modulated acoustic components associated with electric propulsion drives and motor-control electronics. Characterizing these signatures is relevant to passive detection and engineering diagnostics, but remains challenging in coastal environments because ambient noise, shallow-water propagation, and aspect-dependent radiation can obscure vehicle-related features. Existing underwater radiated noise (URN) standards, developed primarily for surface vessels, do not address the spectral, operational, and geometric complexity of SEV measurements. This paper presents an eight-step methodology for SEV URN characterization, covering measurement design, cavitation assessment, frequency-band selection, ambient-noise characterization, spectral and time--frequency analysis, subsystem-oriented interpretation, propagation-corrected source-related estimation, and angular and operational analysis. The novelty lies in integrating calibrated pass-by acoustics with synchronized vehicle metadata, ambient-noise context, and subsystem-oriented analysis to resolve tonal and modulated features that broadband methods cannot capture. The methodology is demonstrated using an A18D AUV measured in coastal water. Drive-related tonal groups were observed near \(5.56\), \(11.1\), and \(22.2~\mathrm{kHz}\), with harmonic structure up to \(105~\mathrm{kHz}\). Source-related tonal PSD estimates ranged from \(77\) to \(120~\mathrm{dB}\) re \(1~\mu\mathrm{Pa}^{2}/\mathrm{Hz}\) at \(1~\mathrm{m}\).
\end{abstract}


\end{frontmatter}

\section{Introduction}
\label{sec:intro}

Submerged electric vehicles (SEVs)---including autonomous underwater vehicles (AUVs), remotely operated vehicles (ROVs), and diver propulsion vehicles (DPVs)---are increasingly used for environmental monitoring, hydrographic surveying, infrastructure inspection, and scientific sampling~\cite{wynn2014auv,hwang2019auv,capocci2017inspection}. The growing recognition of underwater radiated noise (URN) as a systemic environmental challenge~\cite{oceanoise2026} and the emerging requirement that acoustic impact assessments be integrated from the earliest stages of the design and deployment of new ocean technologies underscore the need for standardized, reproducible characterization methods for anthropogenic acoustic sources, including SEVs.

Although their broadband URN is often relatively low because they are driven by compact electric motors, SEVs may produce distinct tonal, harmonic, and modulated components that are relevant to passive detection, classification, localization, and system-state interpretation~\cite{wibisono2023uuv,kita2022tracking,griffiths2001autosub}. From an operational perspective, passive acoustic signatures can support the detection or tracking of SEVs without active transmissions, which is valuable in coastal environments where active systems may be affected by multipath propagation and interference with other acoustic users. The measured acoustic spectrum may provide indirect information on propulsion state, motor-control electronics, rotating components, and possible mechanical or electrical anomalies.

Characterizing SEV URN in coastal environments is challenging. Proximity to ports and marinas can generate dense vessel traffic~\cite{Syrjala2020BalticShipping,garrett2016falmouth}. Elevated biological activity in coastal waters can contribute broadband and impulsive noise~\cite{hildebrand2009ambient,au1998snapping}, while breaking waves and surf near the shore generate broadband noise that dominates the low-frequency coastal soundscape~\cite{haxel2013shallow}. Shallow-water propagation can further modify the received field through repeated surface and seabed interactions, affecting both received level and apparent spectral structure~\cite{katsnelson2002shallow}. Consequently, vehicle-related features may be detectable only within limited frequency bands or during short intervals near the closest point of approach (CPA). Characterization therefore cannot rely on broadband levels alone, but must resolve narrowband and time-varying spectral features that can be distinguished from the ambient background. Close-range measurements, although often necessary to obtain sufficient signal-to-noise ratio, are also sensitive to source--receiver geometry uncertainties and to aspect-dependent radiation, since SEVs cannot generally be assumed to radiate isotropically~\cite{gebbie2012aspect}. In addition, the emitted signature is operationally dependent as tonal components associated with propulsion motors, propeller rotation and pulse-width modulation (PWM)-controlled electric drives, and auxiliary systems~\cite{railey2020propeller,lebesnerais2010pwm} may remain fixed in frequency or vary with speed and motor revolutions per minute (RPM). Meaningful interpretation, therefore, requires acoustic measurements to be accompanied by detailed vehicle metadata such as speed, course, depth, and location.
Existing URN standards, including ISO~17208-1 and ANSI\allowbreak/ASA~S12.64, were developed primarily for underway surface vessels~\cite{iso17208_1_2016,ansiasa_s12_64_2009} and do not address several conditions typical of SEV measurements. In particular, because SEV propulsors often operate at lower tip speeds and greater submergence than conventional surface-vessel propellers, cavitation may be less likely~\cite{ross1987mechanics}, and the radiated signature may be governed more by tonal, harmonic, and modulated components than by broadband cavitation noise. Characterization, therefore, requires resolving the spectral, operational, and directional structure of the radiated signature, including speed dependence, subsystem-related frequencies, and aspect-dependent variation. A measurement and analysis framework that integrates calibrated acoustics, ambient-noise assessment, vehicle metadata, geometry-aware propagation correction, and subsystem-oriented spectral interpretation is therefore needed.
The main parameters used throughout the proposed workflow are summarized in
Table~\ref{tab:key_parameters_intro} to support readability of the subsequent
methodology steps.

\begin{table}[!htbp]
\centering
\footnotesize
\caption{Key parameters used in the proposed SEV underwater-radiated-noise characterization framework.}
\label{tab:key_parameters_intro}
\setlength{\tabcolsep}{4pt}
\renewcommand{\arraystretch}{1.12}
\begin{tabularx}{\textwidth}{@{}
  >{\raggedright\arraybackslash}p{0.17\textwidth}
  >{\raggedright\arraybackslash}p{0.27\textwidth}
  >{\raggedright\arraybackslash}X
@{}}
\toprule
\textbf{Parameter} & \textbf{Meaning} & \textbf{Role in the methodology} \\
\midrule

\(r_{\mathrm{CPA}}\) &
Closest-point-of-approach range &
Defines the source--receiver geometry and affects near-field screening, signal-to-noise ratio, and propagation correction. \\

\addlinespace
\(D_{\mathrm{eff}}\) &
Effective radiating source dimension &
Represents the dominant radiating assembly or structural dimension used to assess near-field limitations. \\

\addlinespace
\(r_{\mathrm{NF}}\) &
Near-field transition distance &
Provides a practical lower-bound distance for applying propagation-corrected source-related estimation. \\

\addlinespace
\(f_{\lambda}\) &
Wavelength-based transition frequency &
Identifies the lower-frequency range where the source--receiver spacing becomes comparable to the acoustic wavelength. \\

\addlinespace
\(f_{\mathrm{co}}\) &
Approximate waveguide cutoff frequency &
Provides a shallow-water propagation constraint used when defining the lower analysis frequency. \\

\addlinespace
\(f_{\min}\), \(f_{\max}\) &
Lower and upper analysis-band limits &
Define the frequency interval used for spectral analysis, subsystem interpretation, and source-related estimation. \\

\addlinespace
\(\alpha\) &
Speed-to-RPM ratio &
Converts logged vehicle speed into shaft-rotation frequency for propulsion and motor-order interpretation. \\

\addlinespace
\(f_r\) &
Shaft-rotation frequency &
Used to predict blade-rate, pole-passing, and speed-dependent modulation features. \\

\addlinespace
\(B\), \(N_p\) &
Blade number and motor pole number &
Used to predict blade-rate harmonics and motor pole-related sideband spacings. \\

\addlinespace
\(f_{\mathrm{PWM}}\) &
Pulse-width-modulation carrier frequency &
Used to select the upper analysis frequency and interpret drive-related tonal components and harmonics. \\

\addlinespace
\(\sigma\) &
Cavitation number &
Screening metric used to assess whether broadband cavitation noise is likely to affect the interpretation. \\

\addlinespace
\(N_{\mathrm{avg}}\) &
Number of Welch averages &
Used to evaluate whether the ambient-noise reference is sufficiently averaged for comparison with vehicle-pass spectra. \\

\addlinespace
\(L_{R,\mathrm{PSD}}\) &
Received tonal PSD level &
Calibrated received spectral level measured at the hydrophone. \\

\addlinespace
\(\mathrm{TL}(f,r,z_s,z_r)\) &
Transmission-loss correction &
Propagation correction used to convert received tonal PSD into a source-related estimate. \\

\addlinespace
\(L_{S,\mathrm{PSD}}\) &
Source-related tonal PSD estimate &
Main quantitative output of the propagation-corrected characterization. \\

\addlinespace
\(\delta L_S\) &
Source-related level uncertainty &
Used to determine whether differences between speeds, receiver aspects, or passages are robust. \\

\bottomrule
\end{tabularx}
\end{table}

This paper presents a practical framework for characterizing SEV URN under realistic coastal field conditions and demonstrates it using an AUV as a representative test case. The contribution is threefold:
\begin{enumerate}
    \item It proposes an in-situ workflow that combines calibrated pass-by measurements, ambient-noise assessment, synchronized navigation data, and propagation-aware source-related estimation.
    \item It introduces subsystem-oriented interpretation of tonal, harmonic, and modulated features using vehicle propulsion and motor-control metadata.
    \item It demonstrates the workflow on an AUV test case, resolving drive-related tonal groups and speed-dependent motor-order modulation over a frequency range extending to \(105~\mathrm{kHz}\).
\end{enumerate}
The acoustic analysis addresses three objectives: source-related characterization through propagation-corrected tonal power spectral density (PSD) estimation; subsystem-oriented interpretation of tonal, harmonic, and modulated features in relation to propulsion and motor-control systems; and angular and operational characterization of signature variability with speed and observation aspect. The framework is used to identify dominant spectral features, distinguish ambient-limited from vehicle-related components, and evaluate aspect-related variability. By resolving components up to \(105~\mathrm{kHz}\), the present work extends the characterized frequency range beyond previous SEV studies, which have generally been limited to below \(5\)--\(16~\mathrm{kHz}\)~\cite{griffiths2001autosub,holmes2010uuvnoise,gebbie2012aspect}, and captures high-frequency tonal components relevant to passive detection and diagnostic interpretation.

\section{Review of Existing Characterization Approaches}
\label{sec:existing_characterization}
Current standardized URN measurement procedures were developed primarily for underway surface vessels. ISO~17208-1 and ANSI/ASA~S12.64 specify pass-by geometry, calibrated receiver requirements, background-noise assessment, and source-level estimation for surface ships~\cite{iso17208_1_2016,ansiasa_s12_64_2009}, but do not address several conditions typical of SEV measurements. These include weak acoustic signatures, short CPA ranges, aspect-dependent radiation, operationally variable source features, and the possibility that non-cavitating electric propulsion may produce tonal, harmonic, and modulated signatures rather than broadband cavitation-dominated noise~\cite{gebbie2012aspect,griffiths2001autosub}. Previous work has addressed parts of this problem: early measurements of the \textit{Autosub} AUV identified signatures at low broadband levels~\cite{griffiths2001autosub}, subsequent studies examined self-noise modeling~\cite{cuschieri2001auv,zimmerman2005decreasing} and low- to mid-frequency emissions~\cite{holmes2010uuvnoise}, and propeller-acoustics literature describes blade-rate tones, tip-vortex contributions, and hydrodynamic excitation relevant to submerged propulsors~\cite{ross1987mechanics,carlton2012marine,brennen1995cavitation}. For SEVs with PWM-controlled drives, electronic components may interact with motor electromagnetic fields and produce discrete tonal components and sidebands governed by pole-passing and pole-pair rates~\cite{lo2000pwm_noise,lebesnerais2010pwm}; however, these mechanisms have received limited attention in SEV acoustic characterization.

Measurements of small remotely operated vehicles (ROVs) have confirmed that compact electrically driven platforms produce detectable acoustic signatures~\cite{cai2010characterization,cai2011passive,buszman2018rovnoise}, while Picardi et al.~\cite{picardi2020minimal} proposed design and reporting criteria for underwater robotic platforms, including acoustic considerations relevant to reproducible characterization. Other studies have exploited tonal components, Doppler shifts, and modulation patterns for passive detection and motion estimation of SEVs~\cite{kita2022tracking,rong2022motion,railey2020propeller,zhang2018auvspeed}. In-situ pass-by studies are particularly relevant: aspect-dependent measurements of a REMUS-100 AUV showed substantial variation with observation direction~\cite{gebbie2012aspect}, experimental and numerical studies have examined UUV radiated noise under controlled conditions~\cite{zhang2024uuvnoise,Yu2020AUVRadiatedNoise,liu2024smalluuv}, and passive detection of underwater scooters has extended the problem to smaller personal SEVs~\cite{diamant2025passive}.

Despite this body of work, SEV acoustic characterization remains methodologically fragmented. Ambient-noise levels are rarely reported alongside vehicle signatures, receiver calibration and pass geometry are inconsistently documented, and comparisons between the URNs of systems remain difficult. In shallow coastal water, normal-mode effects, seabed interaction, and waveguide cutoff can introduce frequency-dependent bias into propagation-corrected estimates that is rarely quantified~\cite{katsnelson2002shallow,jensen2011computational}. Most previous studies have also been limited to frequencies below approximately \(5\)--\(16~\mathrm{kHz}\)~\cite{griffiths2001autosub,holmes2010uuvnoise,gebbie2012aspect}, leaving higher-frequency PWM- and motor-drive-related components largely unexamined. The methodology proposed in Section~\ref{sec:methodological_framework} addresses these gaps by linking calibrated measurement, ambient-noise assessment, synchronized vehicle metadata, propagation-aware source-related estimation, and subsystem-oriented spectral interpretation.

\section{Proposed Methodological Framework}
\label{sec:methodological_framework}
Table~\ref{tab:step_summary} lists the required inputs, main decision criteria, and outputs for each step of the framework.

\begin{table}[!htbp]
\centering
\footnotesize
\caption{Summary of required inputs, main decision criteria, and outputs for each step of the SEV URN characterization framework.}
\label{tab:step_summary}
\setlength{\tabcolsep}{4pt}
\begin{tabularx}{\textwidth}{@{}
  >{\raggedright\arraybackslash}p{0.14\textwidth}
  >{\raggedright\arraybackslash}p{0.22\textwidth}
  >{\raggedright\arraybackslash}p{0.35\textwidth}
  >{\raggedright\arraybackslash}p{0.22\textwidth}@{}}
\toprule
\textbf{Step} & \textbf{Primary inputs} & \textbf{Main criterion / decision} & \textbf{Outputs} \\
\midrule
1.~Measurement design & Site, vehicle navigation, receiver geometry, vehicle metadata & Select CPA range outside the near-field region while maintaining sufficient SNR & Geometry specification; $D_{\mathrm{eff}}$; $r_{\mathrm{NF}}$; $f_{\mathrm{NF}}$; pass-selection criteria \\
\addlinespace
2.~Cavitation assessment & Maximum speed; shallowest operating depth; propulsor diameter; RPM relation & Classify cavitation likelihood using $\sigma$ as a screening metric & Tip speed $V_t$; cavitation number $\sigma$; cavitation classification \\
\addlinespace
3.~Band selection & Recorder bandwidth; CPA range; water depth; expected drive/control frequencies & Define $f_{\min}$ from near-field and propagation limits; define $f_{\max}$ from bandwidth and expected drive harmonics & Analysis band $[f_{\min},f_{\max}]$; excluded components stated explicitly \\
\addlinespace
4.~Ambient-noise assessment & Pre- or post-transit references; AIS and contamination checks & Identify ambient-limited and externally contaminated intervals & Reference PSD; ambient-limited bands; excluded intervals \\
\addlinespace
5.~Spectral analysis & Calibrated vehicle-pass recordings; CPA-centred intervals & Identify fixed, Doppler-shifted, harmonic, modulated, and speed-dependent features & Candidate components; feature classes; Doppler/geometry check \\
\addlinespace
6.~Subsystem interpretation & Vehicle metadata; Step~5 candidate components & Compare observed frequencies and sidebands with expected propulsion, motor, controller, and auxiliary-system frequencies & Predicted and observed frequencies; residuals; attribution confidence \\
\addlinespace
7.~Source-related estimation & High-confidence or tentative components; propagation model; geometry uncertainty & Back-propagate only components sufficiently separated from ambient noise & $L_{S,\mathrm{PSD}}(f)$; uncertainty estimate; excluded components \\
\addlinespace
8.~Angular and operational analysis & Multiple passages, speeds, or receiver aspects & Treat differences as robust only when they exceed combined uncertainty & Aspect asymmetries; speed-dependence classifications; operational trends \\
\bottomrule
\end{tabularx}
\end{table}

\subsection{Scope and Structure}
\label{subsec:scope_structure}

\begin{figure}[!htbp]
\centering
\scriptsize
\resizebox{0.98\linewidth}{!}{%
\begin{tikzpicture}[
  node distance=5mm and 6mm,
  line/.style={-Latex, thick},
  block/.style={draw, rounded corners, align=center, text width=31mm, minimum height=10mm, inner sep=4pt}
]
\node[block] (s1) {Step 1\\\textbf{Measurement design}\\geometry, metadata,\\synchronisation};
\node[block, right=of s1] (s2) {Step 2\\\textbf{Cavitation assessment}\\propulsor operating\\conditions};
\node[block, right=of s2] (s3) {Step 3\\\textbf{Frequency-band selection}\\bandwidth, CPA,\\propagation limits};
\node[block, right=of s3] (s4) {Step 4\\\textbf{Ambient-noise assessment}\\local reference,\\contamination screening};
\node[block, below=14mm of s4] (s5) {Step 5\\\textbf{Spectral analysis}\\tonal identification,\\Doppler check};
\node[block, left=of s5] (s6) {Step 6\\\textbf{Subsystem interpretation}\\motor, propulsor,\\controller features};
\node[block, left=of s6] (s7) {Step 7\\\textbf{Source-related estimation}\\propagation correction,\\uncertainty};
\node[block, left=of s7] (s8) {Step 8\\\textbf{Angular and operational analysis}\\aspect, speed,\\operating state};
\draw[line] (s1) -- (s2);
\draw[line] (s2) -- (s3);
\draw[line] (s3) -- (s4);
\draw[line] (s4.south) -- (s5.north);
\draw[line] (s5) -- (s6);
\draw[line] (s6) -- (s7);
\draw[line] (s7) -- (s8);
\end{tikzpicture}%
}
\caption{Proposed SEV URN characterization framework. Steps~1--4 establish the measurement basis, including survey geometry, cavitation screening, frequency-band definition, and ambient-noise characterization. Steps~5--8 perform the acoustic interpretation, including spectral identification, subsystem-oriented attribution, propagation-corrected source-related estimation, and angular or operational assessment.}
\label{fig:framework_flowchart}
\end{figure}
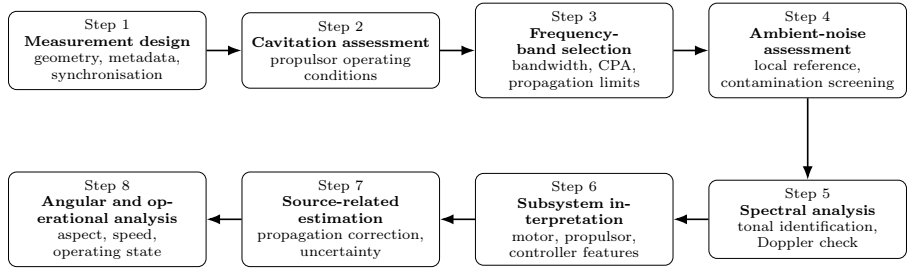

The proposed methodology provides a practical workflow for characterizing URN from SEVs under in-situ coastal conditions. The numerical thresholds used below provide reproducible decision criteria for the present implementation, but they may be adjusted for other SEV platforms, receiver configurations, propagation environments, and signal-to-noise conditions. The workflow consists of eight sequential steps grouped into a measurement phase and an analysis phase. Steps~1--4 establish the measurement basis: pass geometry, cavitation screening, analysis-band selection, and ambient-noise characterization. Steps~5--8 perform the acoustic interpretation: spectral and time--frequency analysis, subsystem-oriented attribution, propagation-corrected source-related estimation, and angular and operational comparison. The workflow is summarized in Fig.~\ref{fig:framework_flowchart}, and the required inputs, main decision criteria, and outputs are listed in Table~\ref{tab:step_summary}. The ordering reflects the dependencies among the operations. Measurement design defines the source--receiver geometry and the validity of later propagation correction. Cavitation screening is then performed before spectral interpretation because broadband cavitation noise would affect the attribution of tonal and motor-drive-related features. Frequency-band selection and ambient-noise characterization define the spectral and environmental context. The analysis phase then identifies candidate components, evaluates their subsystem consistency, estimates source-related tonal PSD levels, and compares angular or operational variability only for components that are sufficiently robust.

The methodology is specifically designed for shallow coastal water, where ambient noise from vessel traffic, biological sources, and breaking waves, combined with multipath propagation and waveguide effects, presents the greatest challenge to resolving vehicle-related tonal components. In deeper or open-ocean environments, some steps --- particularly the waveguide cutoff criterion in Step~3 and the image-source propagation model in Step~7 --- may require replacement with propagation models appropriate to the local bathymetry and water-column structure.

\subsection{Measurement Phase}
\label{subsec:measurement_phase}
The measurement phase defines the experimental basis for the subsequent acoustic analysis. It specifies the pass geometry and required vehicle metadata, screens the operating condition for possible cavitation, selects the usable analysis frequency band, and establishes the local ambient-noise reference. These steps ensure that the later spectral interpretation and source-related estimates are applied only to passages with documented geometry, sufficient signal-to-noise ratio, and a well-defined environmental context.

\subsubsection{Step 1: Measurement Design}
\label{subsec:measurement_design}
The measurement geometry shall be selected to provide sufficient vehicle-to-ambient separation while avoiding conditions in which near-field effects or geometric uncertainty dominate the interpretation. Source-related estimation assumes that the received field can be related to source radiation through a predictable propagation correction. At distances shorter than approximately one acoustic wavelength, or shorter than several characteristic source dimensions, this assumption may not hold and back-propagated estimates become unreliable~\cite{foote2014nearfield_farfield,irs2025urn}.
For compact SEV sources, the effective near-field transition distance is therefore approximated as
\begin{equation}
    r_{\mathrm{NF}} =
    \max\!\left(\frac{c_w}{f},\,3D_{\mathrm{eff}},\,\frac{2D_{\mathrm{eff}}^2 f}{c_w}\right),
    \label{eq:nearfield_distance}
\end{equation}
where $c_w$ is sound speed, $f$ is frequency, and $D_{\mathrm{eff}}$ is the effective radiating dimension of the dominant source assembly. The first term, $\lambda = c_w/f$, is the acoustic wavelength and dominates at low frequencies or short CPA ranges where the receiver lies within one wavelength of the source. The second term, $3D_{\mathrm{eff}}$, provides a geometric near-field margin based on source extent.The third term, \(r_{\mathrm{Ray}} = 2D_{\mathrm{eff}}^{2} f / c_w = 2D_{\mathrm{eff}}^{2}/\lambda\), represents the Rayleigh distance~\cite{foote2014nearfield_farfield}. It becomes the dominant constraint when the effective source dimension is not negligible relative to the acoustic wavelength, such as for larger radiating structures or at higher frequencies where \(D_{\mathrm{eff}}/\lambda\) is no longer small. For SEVs with pulse-width-modulation (PWM)-controlled drives, the upper analysis frequency may extend to tens of kilohertz; in this range, \(r_{\mathrm{Ray}}\) can exceed practical closest-point-of-approach (CPA) ranges even for compact propulsion assemblies. Consequently, source-related estimates at these frequencies should be interpreted as propagation-corrected indicators rather than standardized far-field source levels. The Rayleigh-distance expression is therefore used here as a practical screening criterion, not as a sharp physical boundary.

For propulsion-dominated signatures, \(D_{\mathrm{eff}}\) should represent the motor--shaft--propulsor assembly. Where hull-radiated structural noise is expected to dominate, $D_{\mathrm{eff}}$ should instead represent the relevant vibrating structural dimension, since vibration generated by the motor, shaft, propeller, or drivetrain may be transmitted through the vehicle body and radiated over a larger effective area~\cite{ross1987mechanics,jensen2011computational}. The selected value of $D_{\mathrm{eff}}$ shall therefore be stated explicitly when applying this step.

For a given CPA range \(r_{\mathrm{CPA}}\), the corresponding wavelength-based frequency limit is
\begin{equation}
    f_{\mathrm{NF}} =
    \frac{c_w}{r_{\mathrm{CPA}}}.
    \label{eq:nearfield_frequency_limit}
\end{equation}
Frequencies below this limit may be inspected qualitatively, but they should not be used for propagation-corrected source-related estimation unless far-field conditions can be independently verified.
The CPA range should satisfy the competing requirements of remaining outside the near-field region while maintaining sufficient signal-to-noise ratio (SNR). A practical design window is
\begin{equation}
    \eta r_{\mathrm{NF}} \leq r_{\mathrm{CPA}} \leq r_{\mathrm{max}},
    \qquad 1 < \eta < 3,
    \label{eq:cpa_window}
\end{equation}
where \(\eta\) is a margin factor. The upper range limit may be estimated from
\begin{equation}
    r_{\mathrm{max}} =
    10^{(L_{S,\mathrm{est}}-L_A-\mathrm{SNR}_{\min})/20},
    \label{eq:rmax_snr}
\end{equation}
where \(L_{S,\mathrm{est}}\) is an estimated tonal source level, \(L_A\) is the local ambient PSD level at the same frequency, and \(\mathrm{SNR}_{\min}\) is the minimum required tonal SNR. This expression assumes spherical spreading and is intended only as a conservative pre-deployment design estimate; in shallow coastal water the actual received level may deviate substantially from this prediction due to multipath interference and waveguide effects, and should be refined using a site-appropriate propagation model once environmental data are available. Values of \(\eta\) close to unity maximize SNR but increase sensitivity to geometry errors; larger values reduce near-field risk but may cause weak SEV components to become ambient-limited. If no feasible range satisfies~\eqref{eq:cpa_window}, the survey geometry, site, or operating condition should be revised.

\begin{figure}[!htbp]
\centering
\begin{subfigure}{0.85\textwidth}
    \centering
    \includegraphics[width=0.80\textwidth]{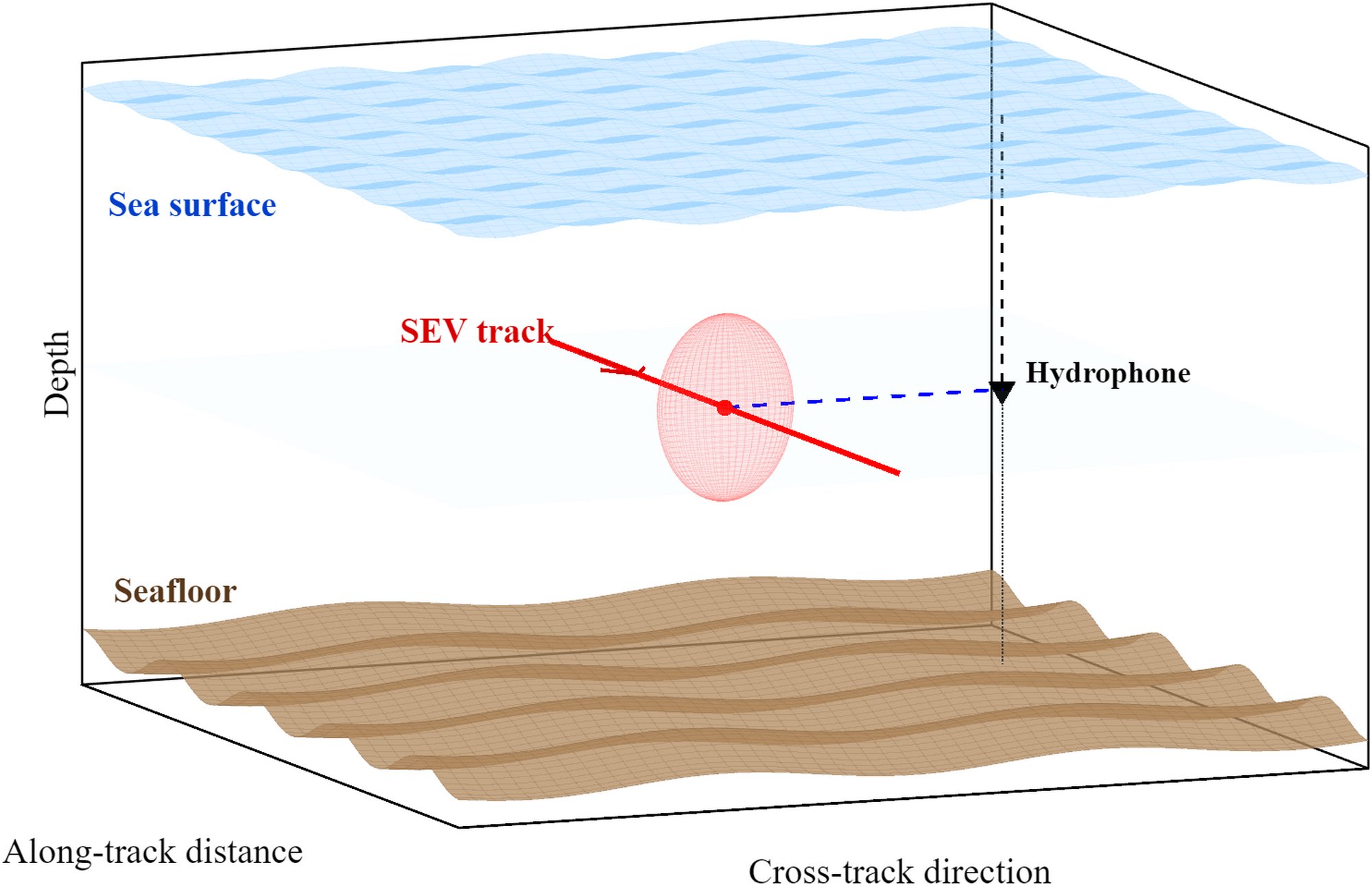}
    \caption{Single-hydrophone configuration. A single receiver is placed near the projected SEV track. The slant closest-point-of-approach distance is denoted \(R_{\mathrm{CPA}}\).}
    \label{fig:sev_geometry_single_hydrophone}
\end{subfigure}
\vspace{0.1cm}
\begin{subfigure}{0.80\textwidth}
    \centering
    \includegraphics[width=0.80\textwidth]{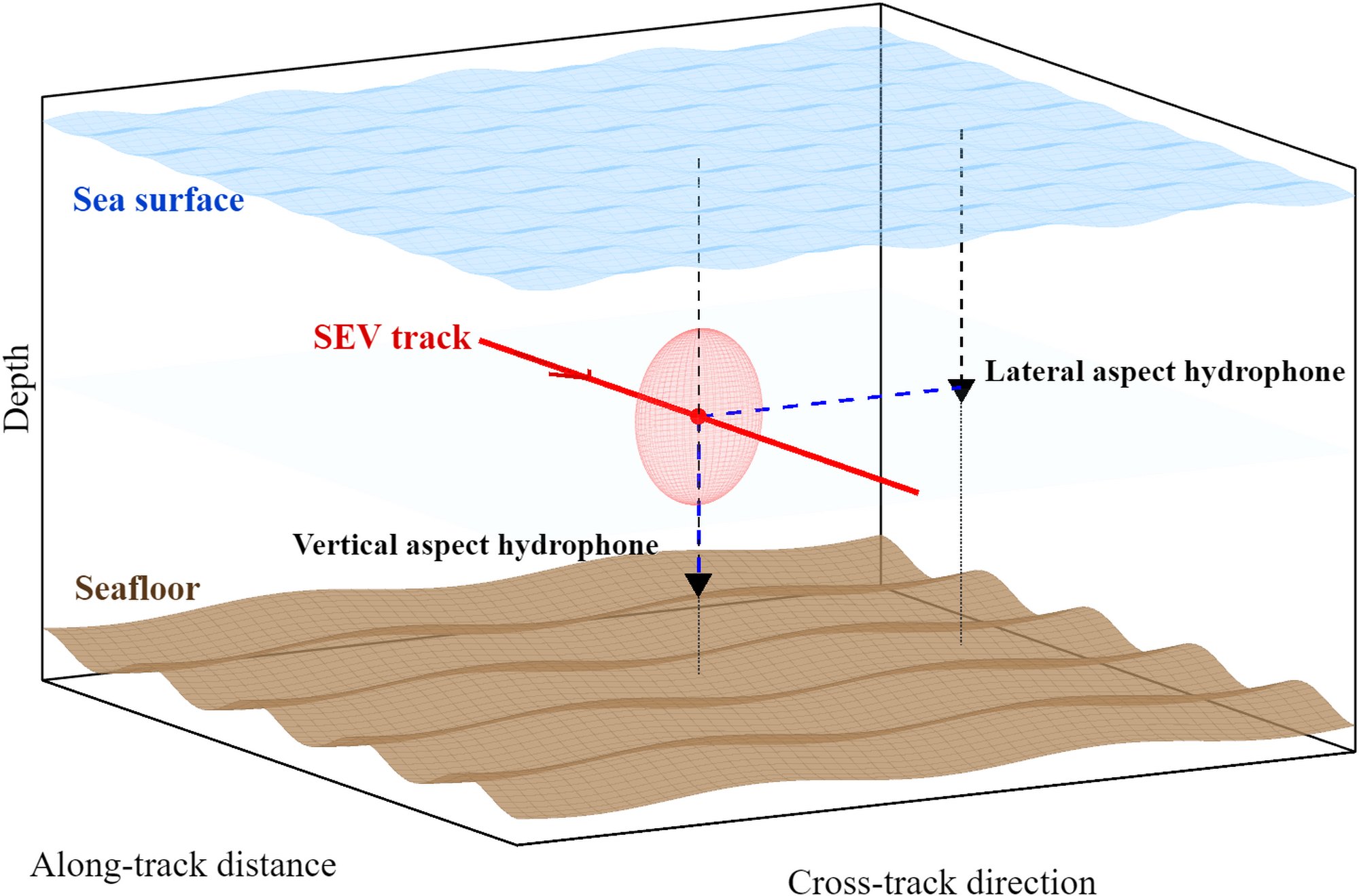}
    \caption{Two-hydrophone configuration. One receiver samples the near-under-track or vertical observation geometry, while a second receiver is placed laterally to sample the cross-track observation geometry during the same run.}
    \label{fig:sev_geometry_two_hydrophones}
\end{subfigure}
\vspace{0.1cm}
\begin{subfigure}{0.80\textwidth}
    \centering
    \includegraphics[width=0.80\textwidth]{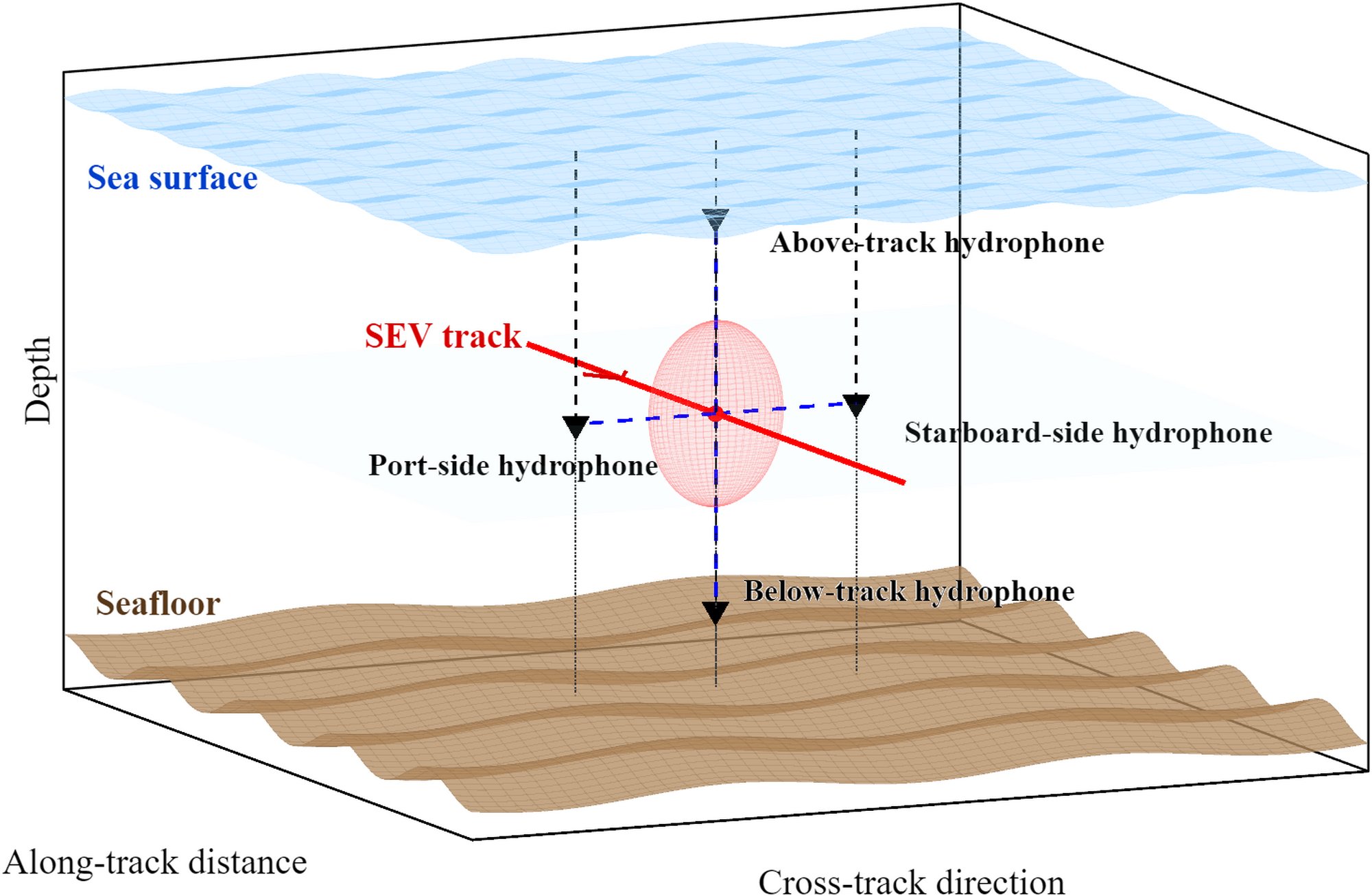}
    \caption{Four-hydrophone configuration. Receivers are distributed above, below, and on both sides of the projected SEV track to provide broader aspect coverage.}
    \label{fig:sev_geometry_four_hydrophones}
\end{subfigure}
\caption{Generic SEV pass-by measurement configurations with increasing geometric coverage: (a)~single hydrophone, (b)~two hydrophones, and (c)~four hydrophones. The red line represents the SEV track, the blue dashed lines indicate \(R_{\mathrm{CPA}}\), and the transparent sphere represents the near-field boundary.}
\label{fig:sev_measurement_geometry_options}
\end{figure}

Where possible, more than one calibrated hydrophone should be deployed during the same run at different cross-track offsets and/or depths. This allows vertical and lateral observation geometries to be compared without relying on separate repeated runs that may differ in speed, depth, heading, or ambient-noise conditions. Three configurations with increasing geometric coverage are illustrated in Fig.~\ref{fig:sev_measurement_geometry_options}: a single hydrophone near the projected track, a two-hydrophone configuration sampling near-under-track and laterally offset aspects, and a four-hydrophone configuration distributed around the track for broader aspect characterization. In all cases, the SEV should follow an approximately controlled pass with documented speed, depth, heading, and timing.

Before the survey, the following vehicle parameters should be documented: propulsor diameter, blade count, shaft speed or speed-to-RPM relationship, motor pole number, PWM frequency and other controller frequencies, auxiliary-system frequencies, commanded speed, operating mode, and active acoustic-system status. This metadata is essential for the subsystem-oriented interpretation in Step~6. Without it, tonal and sideband features can still be detected acoustically, but their physical attribution remains limited.

Passes should be excluded from source-related analysis if the SEV signature is not resolvable above the local ambient reference, since insufficient signal-to-noise ratio prevents reliable tonal identification and renders propagation-corrected source-related estimates meaningless; if external vessel noise is present during the transit or ambient reference interval, since it contaminates both the vehicle-pass recording and the baseline against which vehicle-related components are identified; or if active acoustic transmissions from the vehicle or other sources overlap the analysis band, since these can mask or distort tonal components and introduce spurious spectral features that cannot be distinguished from propulsion- or drive-related signatures. Where transmissions are sparse and time-stamped, affected intervals may be gated out provided the remaining continuous segments are sufficient to resolve the features of interest.

\subsubsection{Step 2: Cavitation Assessment}
\label{subsec:cavitation_assessment}
Cavitation can introduce broadband noise and spectral slopes that may obscure or dominate tonal components~\cite{ross1987mechanics,carlton2012marine,brennen1995cavitation,ittc2021cavitation}, so operating conditions should be screened for possible propulsor cavitation before spectral features are assigned to mechanical, motor-drive, or control-electronic processes. The screening should be performed at the highest vehicle speed and shallowest propulsor submergence used in the analysis, since these conditions maximize propulsor tip speed and minimize hydrostatic pressure, thereby representing the worst-case cavitation risk across all operating conditions considered.
The shaft speed, propulsor tip speed, and cavitation number are
\begin{equation}
    f_r = \frac{\alpha v}{60},
    \qquad
    V_t = \pi D_{\mathrm{prop}} f_r,
    \qquad
    \sigma =
    \frac{p_0+\rho g z-p_v}{0.5\rho V_t^2},
    \label{eq:cavitation_assessment}
\end{equation}
where \(\alpha\) is the speed-to-RPM ratio, \(v\) is vehicle speed, \(D_{\mathrm{prop}}\) is propulsor diameter, \(p_0\) is atmospheric pressure, \(\rho\) is water density, \(g\) is gravitational acceleration, \(z\) is propulsor submergence depth, and \(p_v\) is vapor pressure.

The cavitation number is used here as a practical screening metric. Cavitation inception also depends on blade geometry, loading, inflow non-uniformity, surface condition, and local pressure fluctuations. The following interpretation is therefore used:
\begin{itemize}
    \item \(\sigma > 3\): cavitation is unlikely, and tonal or motor-drive-related interpretation may proceed.
    \item \(1 \leq \sigma \leq 3\): cavitation is possible, and broadband spectral elevation should be interpreted cautiously.
    \item \(\sigma < 1\): cavitation is probable, and broadband components should not be attributed primarily to motor-drive or mechanical sources without additional corroborating evidence.
\end{itemize}

\subsubsection{Step 3: Frequency-Band Selection}
\label{subsec:freq_band_selection}

The analysis frequency band shall be defined before spectral interpretation or source-related estimation is performed. The lower bound should account for both the wavelength-based limit associated with the CPA range and the approximate finite-depth propagation context. The wavelength-based transition frequency is
\begin{equation}
    f_{\lambda} = \frac{c_w}{r_{\mathrm{CPA}}},
    \label{eq:cpa_wavelength_frequency}
\end{equation}
and the approximate shallow-water waveguide cutoff is
\begin{equation}
    f_{\mathrm{co}} = \frac{c_w}{4H},
    \label{eq:waveguide_cutoff}
\end{equation}
where \(H\) is the local water depth. These quantities provide a practical way to avoid using the lowest-frequency range for quantitative source-related estimation when the received field is likely to be strongly affected by finite-depth propagation or wavelength-scale geometry effects~\cite{katsnelson2002shallow,jensen2011computational}.

The lower analysis bound is then selected as
\begin{equation}
    f_{\min} =
    \max\!\left(1.3\,f_{\lambda},\,f_{\mathrm{co}}\right),
    \label{eq:fmin_final}
\end{equation}
or as a more conservative value if local propagation, ambient-noise contamination, or recorder limitations require it. The factor 1.3 provides a practical margin above the wavelength-based transition frequency to account for uncertainty in the reconstructed CPA range and to avoid the immediate transition region where the relative contributions of the direct and reflected paths to the total field are most sensitive to small geometry errors.

The upper bound should be selected from the usable calibrated recorder bandwidth and from the highest drive-, controller-, or auxiliary-system component intended for interpretation. For SEVs with PWM-controlled drives, diagnostically important tonal components may occur near the PWM carrier frequency and its harmonics, and these components can extend above the low- and mid-frequency ranges commonly emphasized in conventional vessel-noise studies. The upper analysis frequency should therefore be selected to include the highest drive-, controller-, or auxiliary-system component retained for interpretation and should satisfy
\begin{equation}
    f_{\max} \geq m f_{\mathrm{PWM}},
    \label{eq:fmax_pwm}
\end{equation}
where \(m\) is the highest PWM harmonic retained and \(f_{\mathrm{PWM}}\) is the PWM carrier frequency. This criterion ensures that the selected band captures the intended carrier frequencies, harmonics, and associated motor-order sidebands without implying that all SEV signatures are necessarily dominated by PWM-related components. The recorder sampling rate must be sufficient to support this upper bound, and the hydrophone and recording chain must remain within their calibrated response over the full analysis band. Any expected drive-related components excluded by the chosen band shall be stated explicitly, so that the scope of the characterization is unambiguous.

\subsubsection{Step 4: Ambient-Noise and Environmental Assessment}
\label{subsec:ambient_environmental_assessment}

A local ambient-noise reference shall be obtained close in time to each SEV transit, using the same hydrophone, gain settings, sampling frequency, calibration procedure, and processing chain as the vehicle-pass recording. This reference is used to identify frequency bands and time intervals in which the SEV signal is separable from the local background.
Using Welch averaging with window duration \(T_w\), reference duration \(T_{\mathrm{ref}}\), and fractional overlap \(\gamma\), the number of averaged spectral segments is
\begin{equation}
    N_{\mathrm{avg}} =
    1 +
    \left\lfloor
    \frac{T_{\mathrm{ref}}-T_w}{T_w(1-\gamma)}
    \right\rfloor .
    \label{eq:number_welch_averages}
\end{equation}
A practical target is \(N_{\mathrm{avg}}\geq100\), although shorter references may be used if stationarity is demonstrated and the limitation is reported~\cite{welch1967use,bendat2010random}. Spectral regions where the vehicle-pass PSD exceeds the ambient reference by less than \(3~\mathrm{dB}\) should be identified as ambient-limited and excluded from source-related estimation.

When AIS or local traffic information is available, nearby vessel contamination should be screened before SEV components are interpreted. A conservative fixed-radius exclusion criterion is recommended: AIS broadcasts should be examined within a predefined radius around the survey area, and intervals during which an AIS-equipped vessel was present inside this radius should be excluded or flagged. The radius should be selected based on local traffic density, the frequency band of interest, and the expected propagation range of vessel noise at the site; the selected radius should be reported explicitly and interpreted as a practical local-screening distance rather than as a strict acoustic detection limit, since vessel noise, especially at low frequencies, may propagate from well beyond any fixed screening radius~\cite{hildebrand2009ambient}. AIS screening should also be combined with direct inspection of the acoustic record for non-AIS vessel noise, active acoustic transmissions, impulsive events, and other non-SEV contamination. Environmental variables relevant to interpretation, including water depth, sound-speed profile, sea state, wind, and biological or impulsive noise conditions, should also be documented.

\subsection{Analysis Phase}
\label{subsec:analysis_phase}

The following steps are applied offline to calibrated passages satisfying the measurement-phase criteria. Multiple passes are advised to increase the statistical range of the comparison.

\subsubsection{Step 5: Spectral and Time--Frequency Analysis}
\label{subsec:spectral_time_frequency_analysis}

Calibrated spectrograms and PSD estimates shall be computed for the ambient reference, the full vehicle passage, and a CPA-centred interval. The analysis window should be long enough to resolve the expected 
minimum sideband spacing \(\Delta f_{\min}\). For a Hann window, the effective noise bandwidth is approximately $2/T_w$, so the window duration should satisfy
\begin{equation}
    T_w \geq \frac{4}{\Delta f_{\min}},
    \label{eq:window_criterion}
\end{equation}
to ensure that adjacent sidebands separated by \(\Delta f_{\min}\) are resolved with a practical margin. The window should nonetheless remain short enough to track level changes near CPA.

Each candidate's spectral feature should be classified as one or more of the following:
\begin{itemize}
    \item fixed: stable in frequency and consistent with an electronically imposed carrier;
    \item Doppler-shifted: varying around CPA in a manner consistent with source--receiver motion;
    \item speed-dependent: center frequency or sideband spacing changes with logged vehicle speed;
    \item harmonic: occurring at an integer multiple of a lower-frequency component;
    \item modulated: sidebands occur around a carrier, with spacing related to shaft, blade, pole-passing, pole-pair, or auxiliary-system rates.
\end{itemize}
Where stable tonal components are present, the Doppler shift can be used as an independent check on the reconstructed pass geometry.
For a straight-line pass at approximately constant speed, the approaching and receding frequencies are related to the source frequency and radial speed by
\begin{equation}
    f_{\mathrm{app}} = f_0\frac{c_w}{c_w-v_r},
    \qquad
    f_{\mathrm{rec}} = f_0\frac{c_w}{c_w+v_r}.
    \label{eq:doppler_forward}
\end{equation}

The corresponding source frequency and radial speed estimates are
\begin{equation}
    f_0 =
    \frac{2f_{\mathrm{app}}f_{\mathrm{rec}}}{f_{\mathrm{app}}+f_{\mathrm{rec}}},
    \qquad
    v_r =
    c_w\frac{f_{\mathrm{app}}-f_0}{f_{\mathrm{app}}}.
    \label{eq:doppler_method}
\end{equation}
The inferred \(v_r\) should be compared with the value predicted from the logged vehicle speed and CPA geometry. Systematic discrepancies should be corrected if justified, or carried forward as part of the geometric uncertainty in Step~7.

\subsubsection{Step 6: Subsystem-Oriented Interpretation}
\label{subsec:subsystem_characterization}

Observed spectral components shall be interpreted by comparing their centre frequencies, harmonic structure, modulation sidebands, recurrence across passages, and speed dependence with frequencies expected from the vehicle propulsion, motor, controller, PWM, and auxiliary systems. This step is not intended to assign every detected tonal component to a physical mechanism. Rather, it provides a structured way to distinguish components that can be linked to documented vehicle metadata from components that remain unidentified. PWM-controlled motor drives may generate discrete tones, harmonics, and modulation sidebands through interactions between switching frequencies, electromagnetic forces, and mechanical response~\cite{lebesnerais2010pwm,lo2000pwm_noise}. The interpretation, therefore, requires the acoustic observations to be evaluated together with the vehicle metadata documented in Step~1.

For each passage, the shaft-rotation frequency is estimated from the logged vehicle speed as
\begin{equation}
    f_r =
    \frac{\alpha v}{60},
    \label{eq:shaft_frequency}
\end{equation}
where \(\alpha\) is the speed-to-RPM ratio and \(v\) is vehicle speed. Frequencies associated with blade-rate forcing and motor pole-passing are then predicted as
\begin{equation}
    f_{B,k} = kBf_r, \qquad k = 1,2,3,\ldots ,
    \label{eq:blade_rate}
\end{equation}
where \(B\) is blade number. If a stable carrier $f_c$ is present, expected motor-order sidebands around that carrier are
\begin{equation}
    f_{c,k} =
    f_c \pm k f_{\mathrm{pole}}
    \quad \,
    \qquad k = 1,2,3,\ldots ,
    \label{eq:sideband_positions}
\end{equation}
where $f_{\mathrm{pole}} = N_p f_r$ is the dominant electromagnetic force frequency, with $N_p$ the number of motor poles. This frequency arises because the fundamental electrical frequency is $f_e = (N_p/2)\,f_r$ (the pole-pair rate), and magnetic force scales as $B^2$, producing a dominant mechanical excitation at $2f_e = N_p f_r$~\cite{lebesnerais2010pwm,lo2000pwm_noise}. The pole-pair rate $f_{\mathrm{pair}} = (N_p/2)\,f_r$ may also appear as a weaker sideband spacing due to asymmetries in the winding or 
drive electronics.

For each observed component or sideband spacing, the residual relative to the nearest predicted frequency is
\begin{equation}
    \epsilon =
    \frac{|f_{\mathrm{obs}}-f_{\mathrm{pred}}|}{f_{\mathrm{pred}}}
    \times 100\%.
    \label{eq:freq_residual}
\end{equation}
The residual is evaluated together with the spectral resolution, recurrence across passages, and speed dependence. In particular, a component whose absolute center frequency remains fixed but whose sideband spacing scales with vehicle speed should be interpreted as a fixed carrier modulated by a speed-dependent mechanical or electromagnetic process, rather than as a simple shaft-order tone.

Each component is assigned one of three attribution levels:
\begin{itemize}
    \item High confidence: the component recurs across at least two independent passages; the observed carrier, harmonic, or sideband spacing agrees with a predicted subsystem frequency within the spectral resolution; and the relevant sideband spacing or center-frequency shift scales with logged vehicle speed or shaft speed as expected.
    \item Tentative: the component recurs or is clearly above ambient noise, and at least one feature agrees with a predicted subsystem frequency or speed-dependent trend, but the physical mechanism is incomplete because recurrence, speed scaling, harmonic relation, or metadata support is insufficient.
    \item Unidentified: the component cannot be related to available subsystem metadata, does not show consistent recurrence, or does not exhibit the expected speed dependence.
\end{itemize}

Only components assigned a high-confidence or tentative attribution should be carried forward to source-related estimation in Step~7. Unidentified components may be reported for diagnostic completeness, but they should not be used as the basis for physical interpretation.

\subsubsection{Step 7: Source-Related Estimation and Propagation Correction}
\label{subsec:source_related_estimation}

Source-related tonal PSD levels shall be estimated only for components that are sufficiently separated from the ambient reference and have at least tentative subsystem attribution. Only tonal components exceeding the local ambient reference were back-propagated. The source-related tonal PSD estimate is
\begin{equation}
    L_{S,\mathrm{PSD}}(f) =
    L_{R,\mathrm{PSD}}(f)
    +
    \mathrm{TL}(f,r,z_s,z_r),
    \label{eq:sl_psd_backprop_method}
\end{equation}
where $L_{S,\mathrm{PSD}}$ is the source-related tonal PSD estimate expressed in dB re $1~\mu\mathrm{Pa}^2/\mathrm{Hz}$ at $1~\mathrm{m}$, $L_{R,\mathrm{PSD}}$ is the received PSD level in dB re $1~\mu\mathrm{Pa}^2/\mathrm{Hz}$, $\mathrm{TL}(f, r, z_s, z_r)$ is the frequency-dependent transmission-loss correction in dB, $r$ is the source--receiver range, $z_s$ is the source depth, and $z_r$ is the receiver depth.
The propagation model should be selected according to range, water depth, bathymetry, and expected boundary interaction~\cite{jensen2011computational,katsnelson2002shallow}. For very short ranges where boundary reflections are negligible, spherical spreading may be sufficient. For short-range shallow-water measurements where direct, surface-reflected, and bottom-reflected paths may interact, an image-source or other boundary-aware model should be used. For longer ranges, range-dependent bathymetry, or strong modal structure, a normal-mode, ray, or parabolic-equation model is more appropriate.
Transmission loss should be averaged consistently over the frequency interval used for each tonal or narrowband estimate rather than evaluated at a single FFT bin. At low frequencies, the TL averaging bandwidth should be informed by the modal spacing of the waveguide; a reference value is
\begin{equation}
    \Delta f_{\mathrm{ref}} = \frac{c_w}{2H}.
    \label{eq:tl_bandwidth}
\end{equation}
Where the band is wider than $\Delta f_{\mathrm{ref}}$, the averaged TL represents a modal energy average rather than a narrow-band modal value; this is acceptable when the propagation geometry is well constrained, and the frequency-dependent TL variability within the band is small relative to the geometric uncertainty $\delta L_S$.

Geometric uncertainty shall be evaluated by recomputing TL for perturbed source--receiver geometry. The range uncertainty associated with a timing error $\delta t$, where $\delta t$ is the uncertainty in the reconstructed time of CPA, is
\begin{equation}
    \delta r = v\,\delta t.
    \label{eq:range_uncertainty}
\end{equation}
The combined geometric uncertainty in the source-related estimate is
\begin{equation}
    \delta L_S =
    \sqrt{(\delta L_r)^2+(\delta L_z)^2},
    \label{eq:geometric_uncertainty}
\end{equation}
where \(\delta L_r\) and \(\delta L_z\) are the TL deviations associated with range and depth uncertainty, respectively. The resulting values should be reported as propagation-corrected source-related estimates, not as standardized absolute source levels, because short-range shallow-water measurements remain sensitive to geometry and boundary conditions.

\subsubsection{Step 8: Angular and Operational Analysis}
\label{subsec:angular_operational_analysis}

Angular and operational variability shall be assessed only after Step~7 has identified components with sufficient ambient separation, subsystem attribution, and uncertainty estimates. A level difference between two source-related estimates is treated as robust only if it exceeds the combined uncertainty of the compared estimates, including both geometric and systematic contributions. For any pairwise comparison between estimates $L_{S,1}$ and $L_{S,2}$,
\begin{equation}
    |\Delta L| >
    k\sqrt{\delta L_{S,1}^{2}+\delta L_{S,2}^{2}},
    \label{eq:aspect_difference_criterion}
\end{equation}
where $k \geq 1$ is a coverage factor. A value of $k=1$ corresponds to the RSS-equivalent threshold appropriate when uncertainties are purely geometric and independent. A more conservative value of $k =\sqrt{2}$ is recommended when systematic contributions --- such as propagation-model uncertainty, residual Doppler offsets, or seabed property uncertainty --- are present but not fully quantified, since these contributions are common to both estimates being compared and therefore do not cancel in the difference. For speed comparisons 
involving a single receiver, $\delta L_{S,1} = \delta L_{S,2} = 
\delta L_S$, so~\eqref{eq:aspect_difference_criterion} reduces to
\begin{equation}
    |\Delta L| > k\sqrt{2}\,\delta L_S,
    \label{eq:speed_dependence_criterion}
\end{equation}
which evaluates to approximately $2\delta L_S$ for $k = \sqrt{2}$, 
consistent with the conservative threshold applied in the test case.

For single-pass aspect analysis, received levels may be compared at symmetric time offsets before and after CPA. Under approximately constant speed and depth, equal time offsets correspond to approximately equal source--receiver ranges, so residual differences are more likely to reflect aspect-dependent radiation or geometry-sensitive propagation than simple spreading loss. The comparison window should satisfy
\begin{equation}
    T_{\mathrm{asp}} \leq \frac{r_{\mathrm{CPA}}}{v}.
    \label{eq:aspect_window}
\end{equation}
Such results should be reported as before--after received-level asymmetries rather than as full source directivity estimates unless the receiver geometry and propagation model are sufficient to separate directivity from propagation effects.

Speed dependence shall be assessed by comparing the same tonal component or modulation pattern across passages with different logged speeds. Where shaft speed or RPM changes during a transit, sideband spacings should be checked against the expected time-varying pole-passing rate,
\begin{equation}
    \Delta f_{\mathrm{pole}}(t) =
    N_p \frac{\alpha v(t)}{60}.
    \label{eq:sideband_evolution}
\end{equation}
The agreement between the observed sideband evolution and~\eqref{eq:sideband_evolution} supports the subsystem attribution from Step~6. Disagreement should lead to downgrading the attribution confidence or reclassifying the component as unidentified.

\section{AUV Test Case}
\label{sec:auv_test_case}

\subsection{AUV Platform}
\label{subsec:auv_platform}

\begin{figure}[!htbp]
\centering
\includegraphics[width=0.8\textwidth]{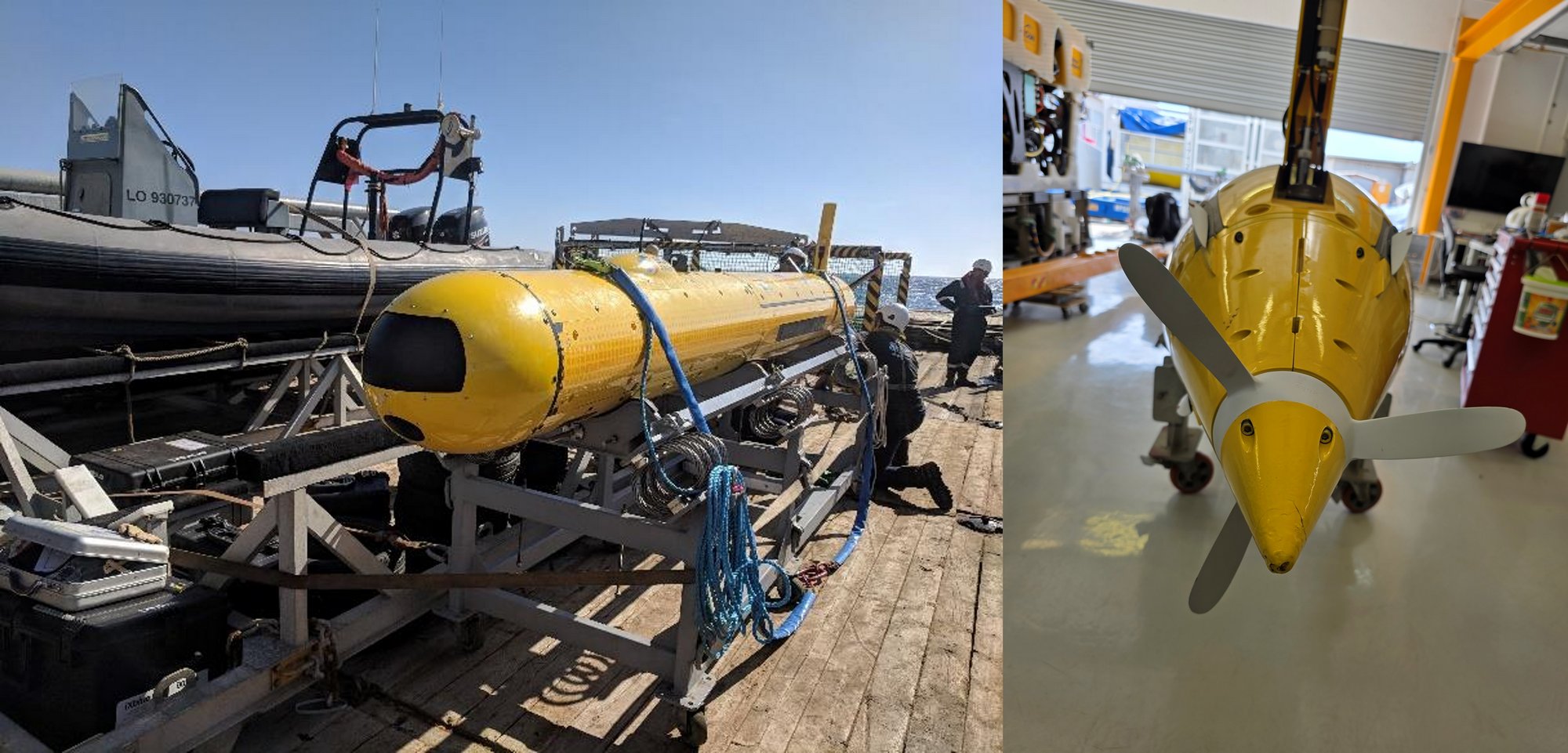}
\caption{A18D AUV used in the experimental test case, including the three-bladed propeller.}
\label{fig:auv_photo}
\end{figure}

The proposed framework was demonstrated using an A18D AUV manufactured by ECA Group~\cite{haifa_a18d_auv} (Fig.~\ref{fig:auv_photo}). The vehicle is approximately $5.5~\mathrm{m}$ long, $0.5~\mathrm{m}$ in diameter, weighs approximately $720~\mathrm{kg}$, and is rated to $3000~\mathrm{m}$ depth. It was selected as a representative SEV test case because it combines 
compact electric propulsion with onboard navigation, communication, 
sensing, and control systems that may contribute to the measured 
acoustic signature. The navigation system includes a Teledyne RDI 
Navigator $1200~\mathrm{kHz}$ DVL~\cite{trdi_navigator_2017} for 
bottom-track velocity estimation.

\subsubsection{Acoustically Relevant Systems}
\label{subsubsec:acoustically_relevant_systems}

The propulsion system consists of a \textit{Megaflux MF150} housed brushless torque motor with 24 poles~\cite{alliedmotion_megaflux_datasheet}, driving a three-bladed composite propeller of approximately $0.4~\mathrm{m}$ diameter. The vehicle supports speeds of approximately $1$--$2.9~\mathrm{m\,s^{-1}}$. Based on the available vehicle metadata, the motor-speed relationship was approximately $180~\mathrm{rpm}$ per $1~\mathrm{m\,s^{-1}}$. The motor is controlled by an \textit{Elmo Guitar} digital servo drive, which has a continuous power rating of up to $4.8~\mathrm{kW}$~\cite{elmo_guitar}. Motor voltage is regulated using a PWM, in which the effective applied voltage is controlled by the rapid switching of the power electronics. In the tested configuration, the PWM carrier frequency was approximately $22~\mathrm{kHz}$. The drive uses a nested servo-control architecture with a current loop operating at approximately $11~\mathrm{kHz}$ and a velocity loop operating at approximately $5.5~\mathrm{kHz}$. The current loop regulates motor current and instantaneous electromagnetic torque, whereas the velocity loop determines the torque demand required to maintain the commanded motor speed. These controller rates are acoustically relevant because PWM-controlled motor drives can generate harmonic current components that interact with motor electromagnetic fields and mechanical resonances, producing discrete tonal components, harmonics, and sidebands~\cite{lo2000pwm_noise,lebesnerais2010pwm}. A summary of the propulsion system and drive parameters relevant to the acoustic interpretation is provided in Table~\ref{tab:auv_drive_specs}.

\begin{table}[!htbp]
\centering
\small
\caption{Propulsion-system and drive characteristics of acoustic relevance for the tested AUV.}
\label{tab:auv_drive_specs}
\begin{tabularx}{\textwidth}{@{}p{0.48\textwidth}X@{}}
\toprule
\textbf{Characteristic} & \textbf{Specification} \\
\midrule
Propeller configuration & Three-bladed $0.4~\mathrm{m}$ diameter propeller \\
Vehicle speed range & Approximately $1$--$2.9~\mathrm{m\,s^{-1}}$ \\
Speed-to-motor-speed ratio & Approximately $180~\mathrm{rpm}$ per $1~\mathrm{m\,s^{-1}}$ \\
Maximum motor rotational speed & Approximately $522~\mathrm{rpm}$ at $2.9~\mathrm{m\,s^{-1}}$ \\
Motor pole number & 24 \\
PWM carrier frequency & Approximately $22~\mathrm{kHz}$ \\
Current-loop update rate & Approximately $11~\mathrm{kHz}$ \\
Velocity-loop update rate & Approximately $5.5~\mathrm{kHz}$ \\
\bottomrule
\end{tabularx}
\end{table}

\subsection{Instrumentation and Survey Setting}
\label{subsec:instrumentation_survey}

\begin{figure}[!htbp]
\centering
\includegraphics[width=0.8\textwidth]{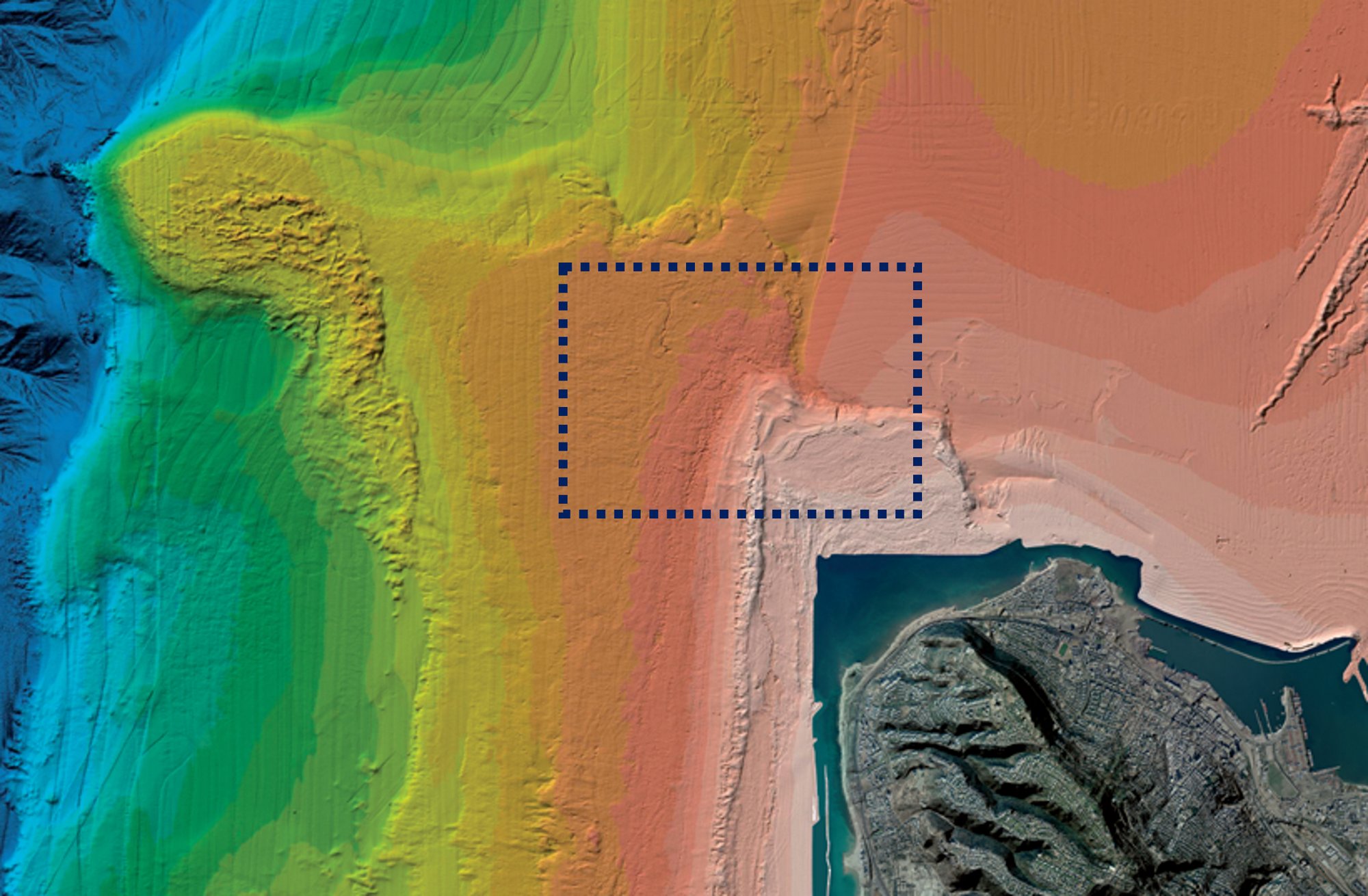}
\caption{Bathymetric map of the Haifa Bay region, including the survey area approximately \(3~\mathrm{nmi}\) northwest of Haifa, Israel. The dashed rectangle marks the survey area. Map source: Geological Survey of Israel~\cite{gsi_haifa_bathymetric_map}.}
\label{fig:survey_area}
\end{figure}

The survey was conducted on 7 January 2026 in a shallow coastal area approximately $3~\mathrm{nmi}$ northwest of Haifa, Israel (Fig.~\ref{fig:survey_area}). The site provided a representative coastal setting for measuring the AUV acoustic signature under operational field conditions, with a mean water depth of approximately $33~\mathrm{m}$.

Two Ocean Sonics icListen HF omnidirectional hydrophones were deployed, each with a sensitivity of $-172~\mathrm{dB}$ re $1~\mathrm{V}/\mu\mathrm{Pa}$, and a sampling rate of $256~\mathrm{kHz}$. The hydrophone coordinates and deployment depths are listed in Table~\ref{tbl:hydrophone_locations}. Consistent with the two-hydrophone configuration recommended in Step~1, the receivers were positioned to provide different source--receiver geometries during the same AUV run: HYD1 was placed close to the projected AUV track to provide a near-under-track geometry suitable for close-range source-related estimation, while HYD2 was laterally offset to provide a simultaneous cross-track observation for aspect comparison. This configuration allowed the same AUV transit to be used for both source-related estimation and geometry-sensitive comparison without relying on separate repeated runs.

\begin{table}[!htbp]
\centering
\small
\caption{Hydrophone coordinates and deployment depths during the AUV survey.}
\label{tbl:hydrophone_locations}
\begin{tabularx}{\textwidth}{@{}lXXc@{}}
\toprule
Hydrophone & Latitude & Longitude & Depth (m) \\
\midrule
Hydrophone 1 (HYD1) & $32.855333^\circ$~N & $34.932383^\circ$~E & 30 \\
Hydrophone 2 (HYD2) & $32.856432236^\circ$~N & $34.933671404^\circ$~E & 15 \\
\bottomrule
\end{tabularx}
\end{table}

\begin{figure}[!htbp]
\centering
\includegraphics[width=0.8\textwidth]{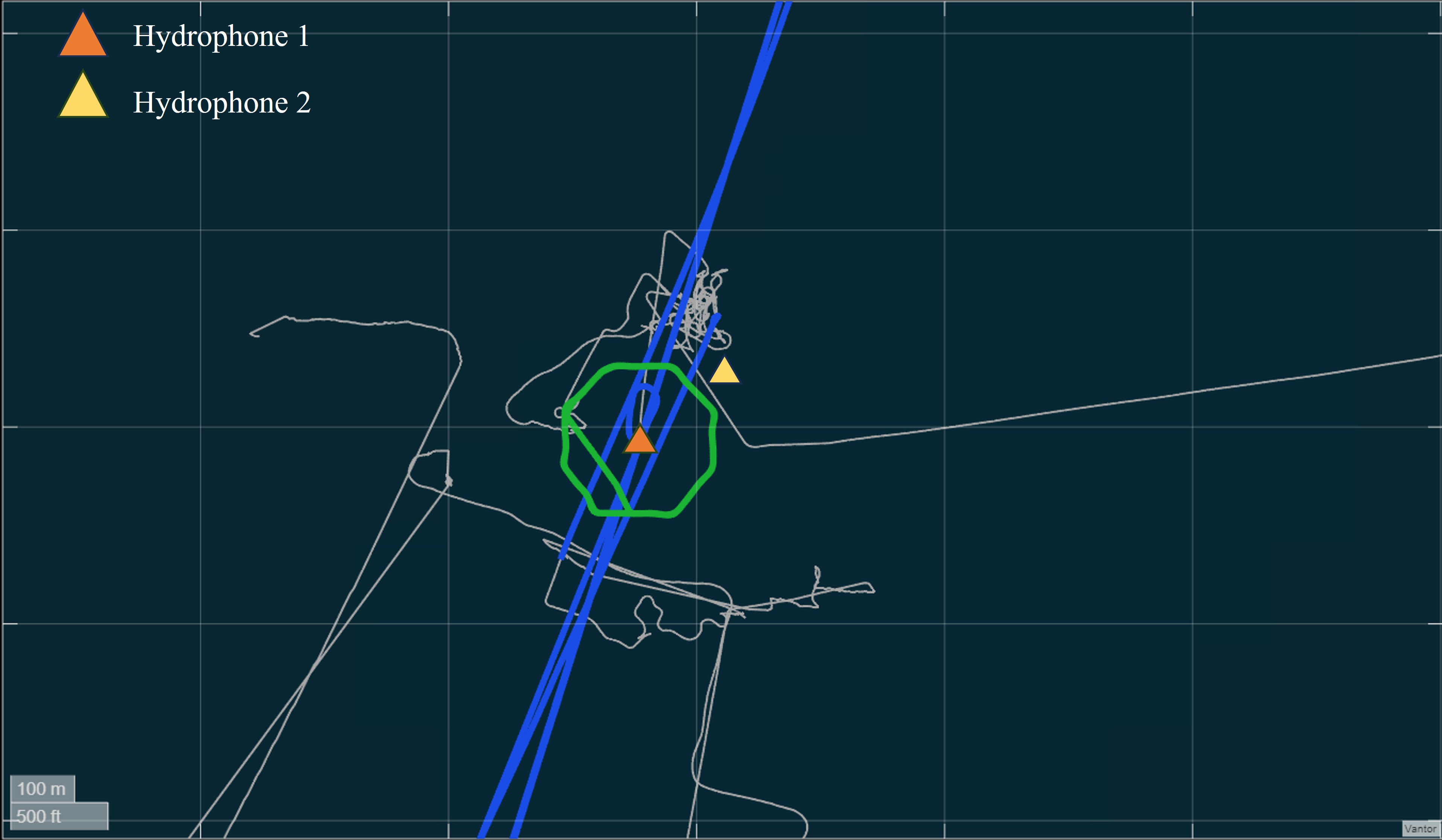}
\caption{Plan view of the AUV track during the survey, showing HYD1 and HYD2 relative to the vehicle trajectory. The blue line indicates the near-under-track passes at a CPA slant range of approximately $10~\mathrm{m}$ from HYD1, corresponding to the passages selected for source-related estimation and aspect comparison. The green polygon indicates additional lateral runs conducted in the vicinity of HYD1. All other transits are shown in grey.}
\label{fig:AUV_track_2d}
\end{figure}

Environmental conditions during the survey were calm, with a wind speed of approximately $4.1~\mathrm{m\,s^{-1}}$ and a wave height of $0$--$0.2~\mathrm{m}$. Temperature and salinity profiles were obtained from the Copernicus Marine Service reanalysis~\cite{copernicus_medsea_phy_reanalysis} and converted to a sound-speed profile using the Mackenzie equation~\cite{mackenzie1981soundspeed}. The resulting profile had a mean sound speed of $1529.32~\mathrm{m\,s^{-1}}$ and limited vertical variability, with a total range of $0.41~\mathrm{m\,s^{-1}}$ over the water column. The AUV track and hydrophone locations are shown in Fig.~\ref{fig:AUV_track_2d}.

\subsection{Application of the Measurement Phase}
\label{subsec:measurement_phase_application}

Steps~1--4 of the framework were applied to the A18D AUV survey prior to acoustic analysis. The following subsections document the decisions and outputs of each step for this test case.

\subsubsection{Step 1: Measurement Design and Pass Geometry}
\label{subsubsec:step1_application}

The near-field transition distance was evaluated using Eq.~\eqref{eq:nearfield_distance}. For the A18D, the dominant compact radiating assembly was taken as the motor--shaft--propulsor unit. The effective radiating dimension was set to \(D_{\mathrm{eff}}=0.4~\mathrm{m}\), corresponding to the propeller diameter and representing the largest dimension of the propulsor assembly. At the closest CPA range of approximately \(10~\mathrm{m}\), the wavelength-based transition frequency from Eq.~\eqref{eq:nearfield_frequency_limit} is \(f_{\mathrm{NF}}=c_w/r_{\mathrm{CPA}}=1529.32/10\approx153~\mathrm{Hz}\). The adopted lower analysis bound of \(200~\mathrm{Hz}\) therefore provides a conservative margin above the wavelength-based transition associated with the closest source--receiver geometry. At the highest analyzed frequencies, however, the Rayleigh-distance term in Eq.~\eqref{eq:nearfield_distance} may become non-negligible for the assumed \(D_{\mathrm{eff}}=0.4~\mathrm{m}\). The near-field criterion is therefore used here primarily to define the lower-frequency limit for quantitative source-related estimation, while high-frequency source-related estimates from the closest passages are interpreted with additional caution. The sensitivity of this assessment to the assumed effective radiating dimension is discussed in~\ref{app:supp_freq_band}.

The AUV completed 15 measurement transits in proximity to the hydrophones, with CPA slant ranges spanning approximately $10$--$127~\mathrm{m}$. The transits included variations in speed, range, and source--receiver geometry. The Step~1 pass-selection criteria were applied to all 15 transits, and passages were excluded from source-related analysis if any of the following conditions were met: the AUV signature was not resolvable above the ambient-noise reference by at least $5~\mathrm{dB}$ at any identified tonal frequency; external vessel noise was present during the transit or the ambient reference interval; or active acoustic communication signals from the AUV overlapped the analysis band. The survey was designed to provide an initial assessment of geometry-related variability, with HYD1 and HYD2 providing simultaneous near-under-track and laterally offset observations, respectively, consistent with the two-hydrophone configuration of Step~1.

Five of the fifteen transits satisfied all Step~1 and Step~4 criteria for source-related analysis. The remaining transits were excluded due to insufficient ambient separation, external vessel noise, or active acoustic transmissions for the modem. DVL transmissions were identified as periodic broadband impulsive transients recurring at approximately $1~\mathrm{Hz}$, consistent with the DVL ping rate; although the $1.2~\mathrm{MHz}$ acoustic carrier lies above the recording Nyquist limit, each transmission produced a detectable broadband mechanical impulse within the analysis band at close range, visible as short broadband transients. These transients were present only during close-range AUV passages and were absent from the ambient reference intervals, confirming that the reference recordings were unaffected. Affected samples were excised prior to spectral analysis; the excised intervals totaled less than $1\%$ of each passage duration; where individual transients fell within the CPA-centered analysis window, the short duration of each excised sample ($\sim8~\mathrm{ms}$) did not materially affect the Welch-averaged spectral estimates. The remaining continuous segments were sufficient to support the spectral comparisons and source-related estimates reported in Steps~5--7. The five passages selected for the main analysis are summarized in Table~\ref{tab:auv_passages_main}; the full set of identified transits and their geometry is available in the data repository.

\begin{table}[!htbp]
\centering
\small
\caption{Representative AUV passages selected for the main acoustic analysis. The geometry column denotes the approximate source--receiver configuration at CPA. The full set of identified passages and associated metadata is available 
in the data repository.}
\label{tab:auv_passages_main}
\begin{tabular}{c c c c c}
\toprule
\shortstack{Passage\\ID} &
Hydrophone &
\shortstack{Speed\\(m\,s$^{-1}$)} &
\shortstack{Approx.\\geometry} &
\shortstack{CPA slant\\range\\(m)} \\
\midrule
1  & HYD2 & 2.8 & Laterally offset   & 78.31 \\
2  & HYD1 & 2.8 & Near-under-track   & 10.07 \\
3  & HYD1 & 1.8 & Near-under-track   & 10.03 \\
4  & HYD2 & 1.8 & Laterally offset   & 79.47 \\
12$^{\dagger}$ & HYD1 & 1.1 & Near-under-track   & 12.99 \\
\bottomrule
\end{tabular}
\par\smallskip
\raggedright\small $^{\dagger}$Excluded from source-related estimation (Step~7); used for subsystem interpretation (Step~6) only.
\end{table}
Passage~12 involved a turning maneuver from bow to stern at $1.1~\mathrm{m\,s^{-1}}$ and is included for subsystem interpretation in Step~6, specifically to examine the evolution of sideband spacing during successive RPM changes. Consistent with the Step~1 requirement for controlled straight-line passes, Passage~12 was not used for source-related estimation in Step~7.

\subsubsection{Step 2: Cavitation Assessment}
\label{subsubsec:step2_application}

Following the screening procedure of Step~2, the cavitation number was evaluated for the highest operating speed considered in the experiment. At $2.9~\mathrm{m\,s^{-1}}$, the shaft speed was approximately $522~\mathrm{rpm}$, or $8.7~\mathrm{rev\,s^{-1}}$, based on the speed-to-motor-speed relationship in Table~\ref{tab:auv_drive_specs}. With a propeller diameter of $D_{\mathrm{prop}} = 0.40~\mathrm{m}$, the corresponding tip speed from ~\eqref{eq:cavitation_assessment} is
\begin{equation}
V_t = \pi D_{\mathrm{prop}}\, f_r \approx \pi \times 0.40 \times 8.7 \approx 10.9~\mathrm{m\,s^{-1}}.
\label{eq:auv_tip_speed}
\end{equation}
Using $p_0 = 101{,}325~\mathrm{Pa}$, $\rho = 1025~\mathrm{kg\,m^{-3}}$, $g = 9.81~\mathrm{m\,s^{-2}}$, a nominal propeller submergence depth of $z = 20~\mathrm{m}$, and a vapor pressure of $p_v \approx 2100~\mathrm{Pa}$ at approximately $18\,^{\circ}\mathrm{C}$,~\eqref{eq:cavitation_assessment} gives $\sigma \approx 4.9$. Applying the Step~2 decision rule, this value exceeds the threshold of $\sigma > 3$, so cavitation was classified as unlikely during the analyzed measurement runs. The measured signature was therefore interpreted as sub-cavitation propulsion and motor-drive-related radiated noise, and tonal and motor-drive-related interpretation was permitted to proceed. At the lower operating speeds of $1.1$ and $1.8~\mathrm{m\,s^{-1}}$, the reduced tip speed yields higher cavitation numbers, further supporting this classification.

\subsubsection{Step 3: Frequency-Band Selection}
\label{subsubsec:step3_application}

The analysis frequency band was defined by applying Step~3 to the A18D drive parameters in Table~\ref{tab:auv_drive_specs} and the survey geometry established in Step~1.
For the upper bound, the PWM carrier frequency is approximately $22~\mathrm{kHz}$. To satisfy~\eqref{eq:fmax_pwm} for harmonics up to order $m = 4$, the upper bound must satisfy $f_{\max} \geq 4 \times 22 = 88~\mathrm{kHz}$. The icListen HF system operated at a sampling rate of $256~\mathrm{kHz}$, giving a Nyquist limit of $128~\mathrm{kHz}$, which exceeds the required upper bound of $88~\mathrm{kHz}$. The upper analysis limit was therefore set to $f_{\max} = 105~\mathrm{kHz}$, retaining a margin below the Nyquist frequency while capturing the relevant PWM harmonics and their motor-order sidebands within the calibrated recording-chain response. As a check against the near-field criterion of Step~1, the Rayleigh distance at this upper bound is $r_{\mathrm{Ray}} = 2D_{\mathrm{eff}}^2 f_{\max}/c_w = 2\times0.40^2\times105000/1529.32 \approx 21.9~\mathrm{m}$, which exceeds the $10~\mathrm{m}$ CPA range. The frequency at which $r_{\mathrm{Ray}}$ first equals $r_{\mathrm{CPA}}$ is $f_{\mathrm{Ray}} = r_{\mathrm{CPA}}\,c_w/(2D_{\mathrm{eff}}^2) = 10\times1529.32/(2\times0.40^2) \approx 47~\mathrm{kHz}$; source-related estimates above this frequency are subject to an additional near-field caveat and are carried forward with that qualification noted in Step~7.

For the lower bound, the wavelength-based transition frequency was computed from~\eqref{eq:cpa_wavelength_frequency} using the closest passage CPA range of $r_{\mathrm{CPA}} \approx 10~\mathrm{m}$ and mean sound speed $c_w = 1529.32~\mathrm{m\,s^{-1}}$:
\begin{equation}
    f_{\lambda} = \frac{c_w}{R_{\mathrm{CPA}}} = \frac{1529.32}{10} \approx 153~\mathrm{Hz}.
    \label{eq:auv_flambda}
\end{equation}
The waveguide cutoff frequency was estimated from~\eqref{eq:waveguide_cutoff} using the mean water depth of $H = 33~\mathrm{m}$:
\begin{equation}
    f_{\mathrm{co}} = \frac{c_w}{4H} = \frac{1529.32}{4 \times 33} \approx 11.6~\mathrm{Hz}.
    \label{eq:auv_fco}
\end{equation}
Applying~\eqref{eq:fmin_final} gives $f_{\min} = \max(1.3 \times 153,\; 11.6) \approx 199~\mathrm{Hz}$. The lower analysis bound was therefore set to $f_{\min} = 200~\mathrm{Hz}$, providing a conservative margin above the wavelength-based transition frequency and avoiding the lowest-frequency range where shallow-water modal structure, seabed interaction, and waveguide effects may introduce larger uncertainty in propagation-corrected estimates. Frequencies below $200~\mathrm{Hz}$ were not used for source-related estimation but may be inspected qualitatively for diagnostic purposes. Additional details on the sensitivity of the lower bound to the assumed source dimension are provided in~\ref{app:supp_freq_band}.
The principal analysis band was therefore $200~\mathrm{Hz}$--$105~\mathrm{kHz}$. All drive-related components expected from Table~\ref{tab:auv_drive_specs}, including the velocity-loop carrier at $5.5~\mathrm{kHz}$, the current-loop carrier at $11~\mathrm{kHz}$, and the PWM carrier at $22~\mathrm{kHz}$ together with their harmonics and motor-order sidebands, fall within this band and satisfy~\eqref{eq:fmax_pwm}.

\subsubsection{Step 4: Ambient-Noise Characterization}
\label{subsubsec:step4_application}

Applying Step~4, a passage-specific ambient reference was obtained from a $60~\mathrm{s}$ recording made approximately $15~\mathrm{min}$ before each AUV transit, using the same hydrophones, gain settings, sampling frequency, calibration procedure, and processing chain applied to the vehicle-pass measurements. The reference duration was evaluated against the Step~4 criterion using ~\eqref{eq:number_welch_averages}. With an analysis window duration of $T_w = 1~\mathrm{s}$ and $50\%$ overlap, the number of Welch-averaged spectral segments is
\begin{equation}
    N_{\mathrm{avg}}
    =
    1 +
    \left\lfloor
    \frac{60 - 1}{1(1 - 0.5)}
    \right\rfloor
    = 119,
    \label{eq:ambient_reference_averages}
\end{equation}
This satisfies the Step~4 averaging requirement defined in~\eqref{eq:number_welch_averages}, with \(N_{\mathrm{avg}} \geq K = 100\).

Both the ambient reference and vehicle-pass intervals were screened for external vessel noise, active acoustic transmissions, and impulsive contamination in accordance with Step~4. AIS broadcasts were examined for transiting vessels within a $20~\mathrm{km}$ radius of the survey area, applying the fixed-radius exclusion criterion described in Step~4. This radius was selected as a conservative local-screening distance consistent with previous hydrophone--AIS studies~\cite{Haver2021LargeVessel,Syrjala2020BalticShipping} rather than as a strict acoustic detection limit, and acoustic data recorded during periods in which an AIS-equipped vessel was identified within this radius were excluded from analysis\footnote{The $20~\mathrm{km}$ radius was selected as a practical local-screening distance appropriate to the survey area geometry and the moderate vessel traffic density of the eastern Mediterranean coastal zone. Haver et al.~\cite{Haver2021LargeVessel} and Syrj\"{a}l\"{a} et al.~\cite{Syrjala2020BalticShipping} use $20~\mathrm{km}$ as a geographic study-area boundary for vessel traffic characterization rather than as an acoustic exclusion criterion; the present usage adopts the same numerical value for consistency but on a different empirical basis. Because low-frequency shipping noise propagates over ranges far exceeding $20~\mathrm{km}$~\cite{hildebrand2009ambient}, the threshold is applied conservatively and all retained intervals were additionally verified by direct inspection of the acoustic record.}. Spectral regions where the vehicle-pass PSD exceeded the ambient reference by less than $3~\mathrm{dB}$ were identified as ambient-limited and excluded from source-related estimation in Step~7.

Environmental conditions during the survey were calm, with a wind speed of approximately $4.1~\mathrm{m\,s^{-1}}$ and a wave height of $0$--$0.2~\mathrm{m}$. The sound-speed profile had a mean value of $1529.32~\mathrm{m\,s^{-1}}$ and limited vertical variability, with a total range of $0.41~\mathrm{m\,s^{-1}}$ over the water column. These environmental parameters were used to define the local propagation context and to support the source-related interpretation in Step~7.

\subsection{Application of the Analysis Phase}
\label{subsec:analysis_phase_application}

Steps~5--8 were applied to the five passages satisfying the measurement-phase criteria. The following subsections document the outputs of each step for this test case.

\subsubsection{Steps 5 and 6: Spectral Characteristics and Subsystem Association}
\label{subsubsec:spectral_characteristics}

\begin{figure}[!htbp]
\centering
\includegraphics[width=0.9\textwidth]{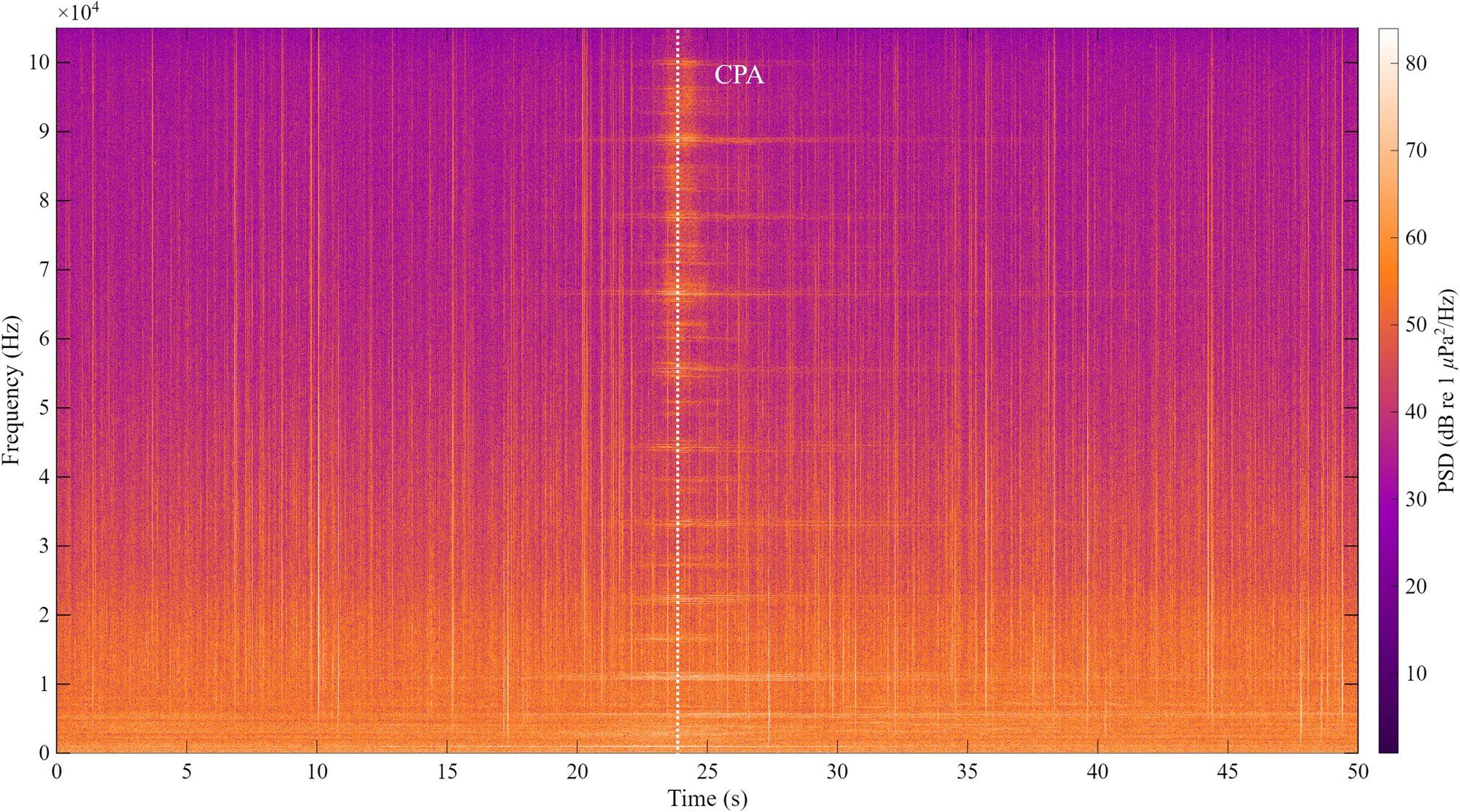}
\caption{Spectrogram of Passage ID~2 recorded by HYD1 over the analyzed frequency band. The white dashed vertical line indicates the CPA time inferred from the Doppler pattern.}
\label{fig:passby_spectrogram}
\end{figure}

Applying Steps~5 and~6, time--frequency analysis was performed on calibrated acoustic recordings from both hydrophones over the selected passage intervals, and candidate tonal components were identified and compared with the predicted subsystem frequencies derived from Table~\ref{tab:auv_drive_specs}. Particular attention was given to Passage IDs~2 and~3, in which the AUV passed HYD1 at an approximate CPA slant range of $10~\mathrm{m}$ at two speeds, $2.8~\mathrm{m\,s^{-1}}$ and $1.8~\mathrm{m\,s^{-1}}$, respectively. These passages provided the clearest basis for comparing the stability and speed dependence of the dominant spectral features. An overview spectrogram for Passage ID~2 over the full analyzed band is shown in Fig.~\ref{fig:passby_spectrogram}.

The received AUV signature was dominated by discrete tonal, harmonic, and modulated components rather than by a broadband elevation of the received spectrum. Vehicle-related components were most clearly observed near CPA, where the vehicle-to-ambient separation was highest. The spectrograms also showed a pronounced ambient-noise background with intermittent transient events, including possible biological contributions such as snapping shrimp. Superimposed on this background, multiple tonal components were observed between approximately $1$ and $100~\mathrm{kHz}$, with a similar overall structure in Passage ID~3 despite the lower vehicle speed. Some components exhibited fluctuating or rippled time--frequency structure, consistent with modulated or mechanically influenced processes, whereas others remained nearly fixed in frequency, consistent with electronically imposed drive-control frequencies. The Doppler-based geometry check of Step~5 was performed using a stable tonal component near $88.7~\mathrm{kHz}$, which is consistent with the fourth harmonic of the $22.2~\mathrm{kHz}$ PWM carrier and its associated motor-order sidebands; the analysis and residual discussion are provided in~\ref{app:doppler}.

\begin{figure}[!htbp]
\centering
\begin{subfigure}[t]{0.48\textwidth}
    \centering
    \includegraphics[width=\textwidth, height=4.5cm, keepaspectratio=false]{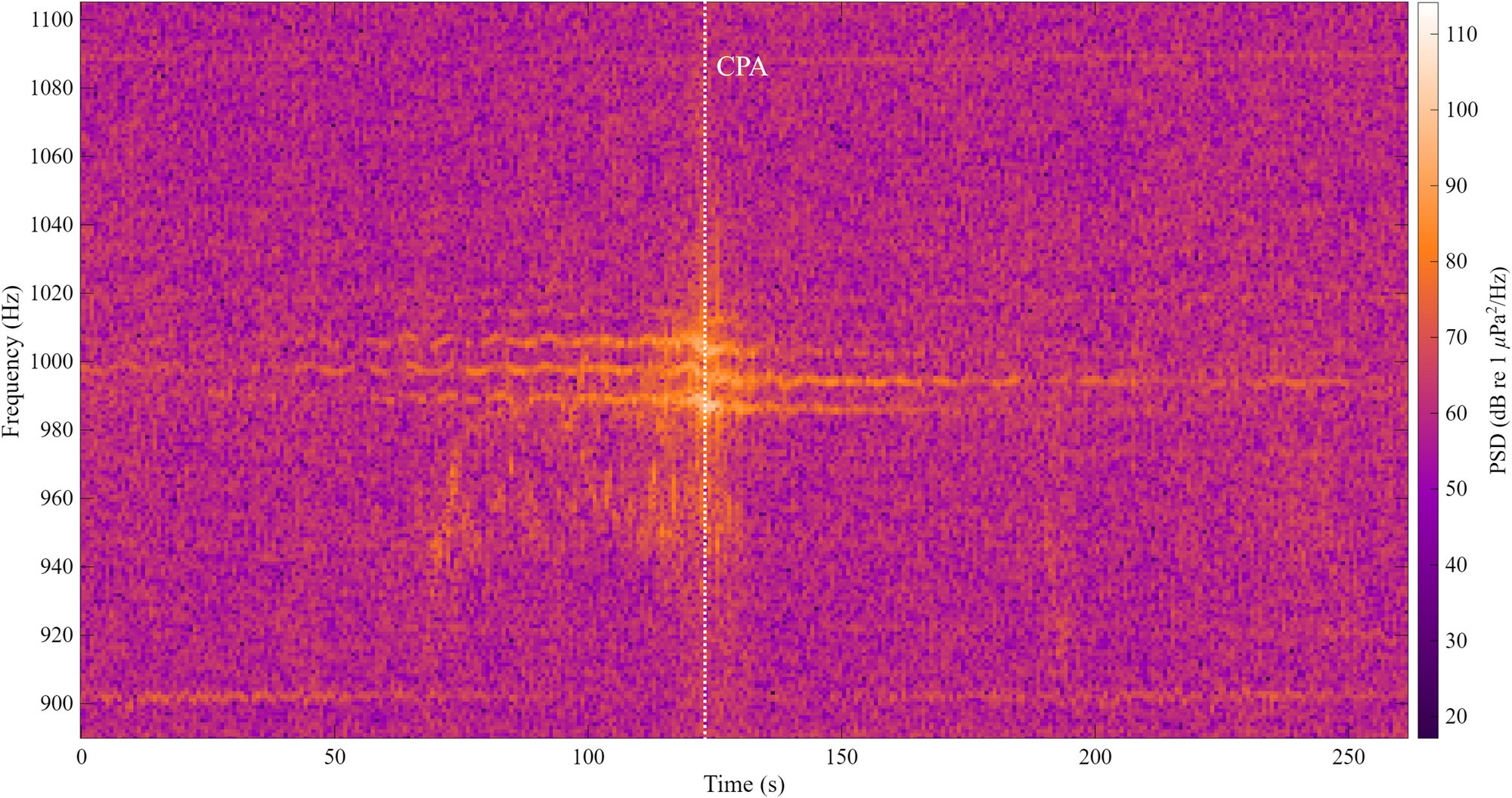}
    \caption{$890$--$1105~\mathrm{Hz}$ band: $\sim\!1~\mathrm{kHz}$ tonal component with shaft-rate modulation spacing.}
    \label{fig:passby_spectrogram_1000}
\end{subfigure}
\hfill
\begin{subfigure}[t]{0.48\textwidth}
    \centering
    \includegraphics[width=\textwidth, height=4.5cm, keepaspectratio=false]{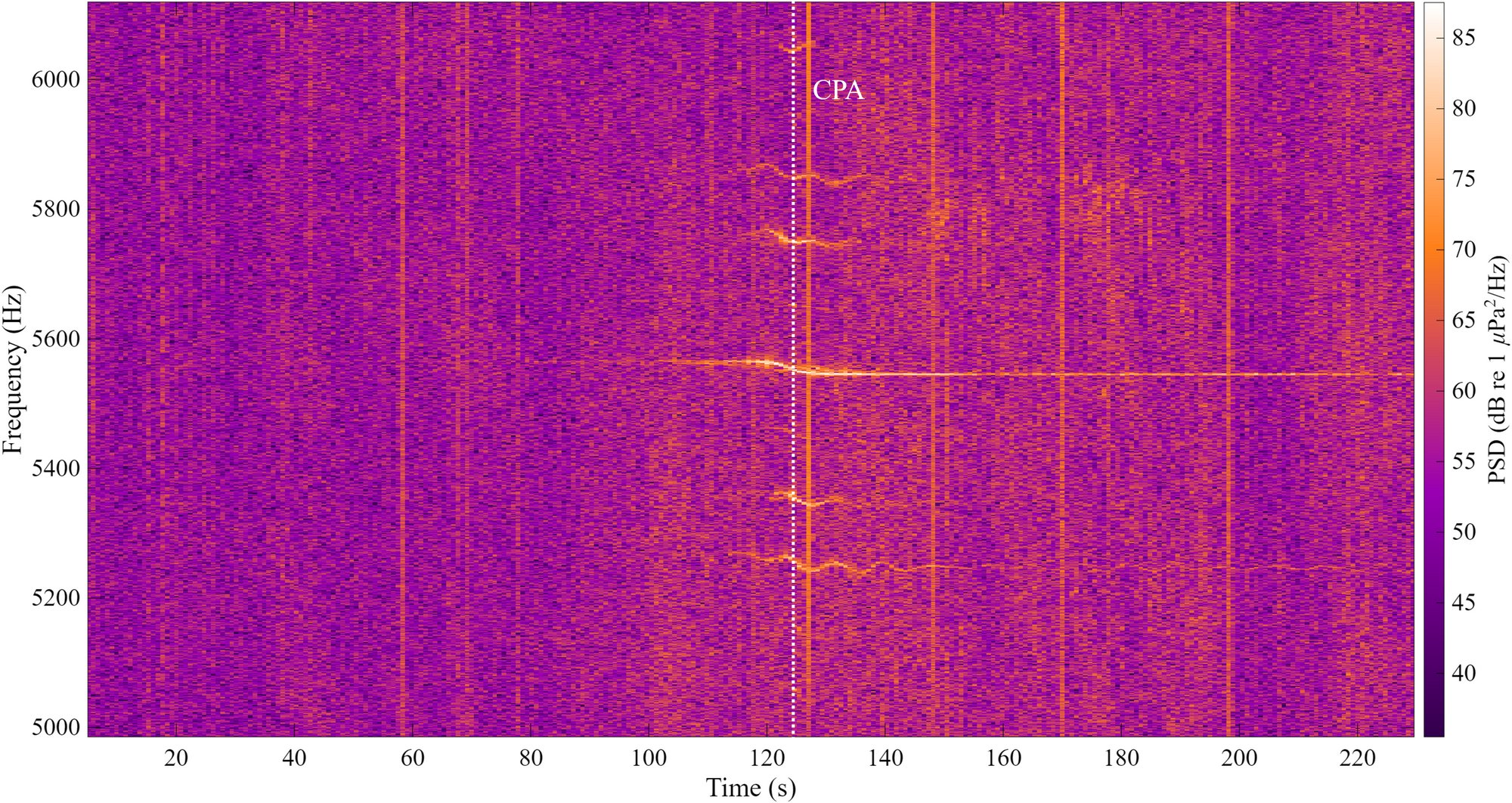}
    \caption{$5000$--$6150~\mathrm{Hz}$ band: velocity-loop carrier at $5556~\mathrm{Hz}$ with pole-passing and pole-pair sidebands.}
    \label{fig:passby_spectrogram_5500}
\end{subfigure}
\vspace{0.3cm}
\begin{subfigure}[t]{0.48\textwidth}
    \centering
    \includegraphics[width=\textwidth, height=4.5cm, keepaspectratio=false]{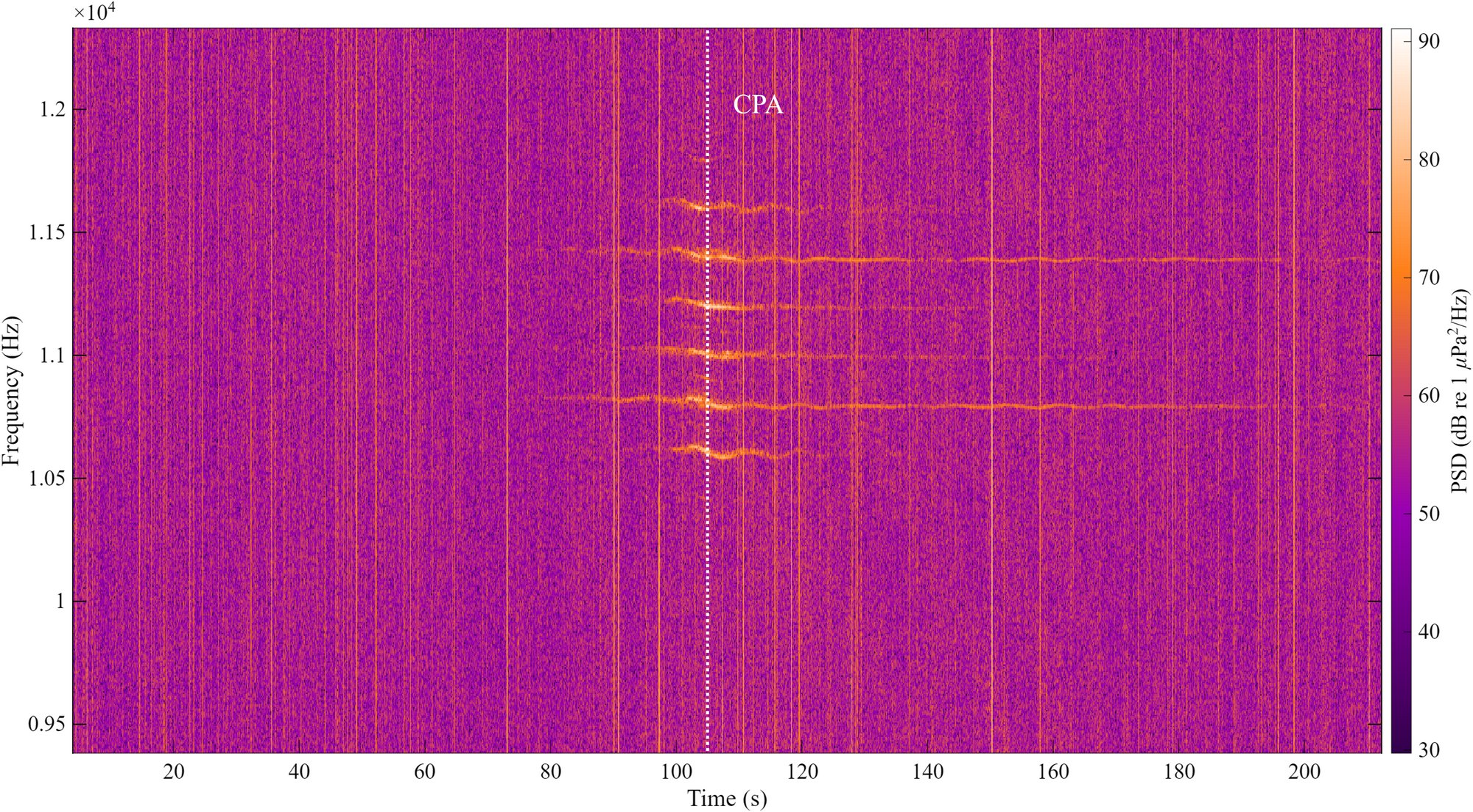}
    \caption{$9380$--$12330~\mathrm{Hz}$ band: current-loop carrier at $11112~\mathrm{Hz}$ with the same sideband structure.}
    \label{fig:passby_spectrogram_11000}
\end{subfigure}
\hfill
\begin{subfigure}[t]{0.48\textwidth}
    \centering
    \includegraphics[width=\textwidth, height=4.5cm, keepaspectratio=false]{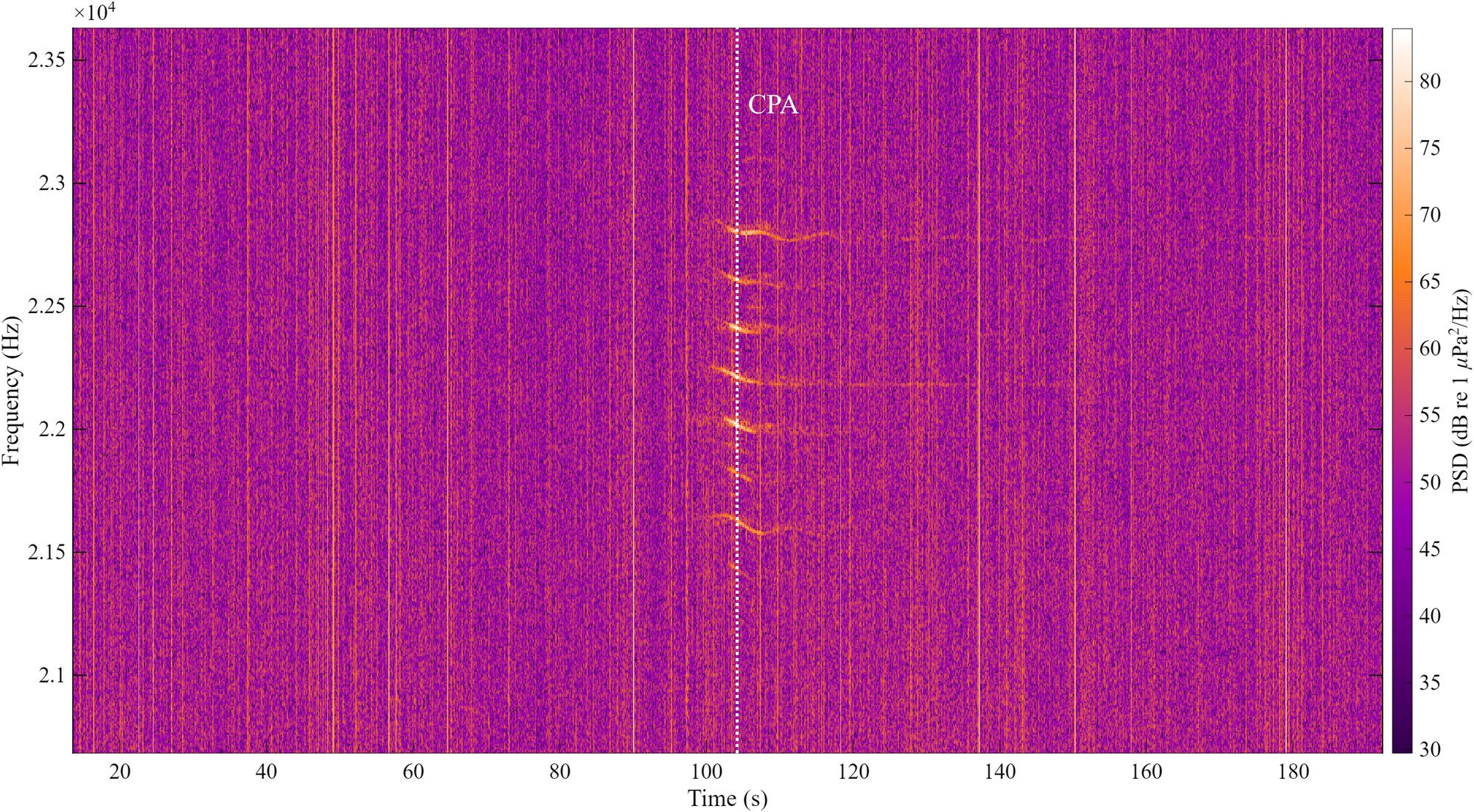}
    \caption{$20680$--$23630~\mathrm{Hz}$ band: PWM carrier at $22224~\mathrm{Hz}$ with speed-dependent motor-order sidebands.}
    \label{fig:passby_spectrogram_22000}
\end{subfigure}
\caption{Spectrograms of Passage ID~2 in the four principal tonal regions (256k FFT, Hann window, 50\% overlap). Each panel shows the CPA time as a white dashed vertical line. Tonal labels correspond to the Step~6 interpretation: A~---~$\sim\!1~\mathrm{kHz}$ tonal, B~---~$5.5~\mathrm{kHz}$ velocity-loop series, C~---~$11~\mathrm{kHz}$ current-loop series, D~---~$22~\mathrm{kHz}$ PWM series.}
\label{fig:passby_spectrograms_tonal}
\end{figure}

At lower frequencies, a clear concentration of tonal energy was observed near $1~\mathrm{kHz}$ (Fig.~\ref{fig:passby_spectrogram_1000}). For Passage ID~2, the dominant component was centered near $997~\mathrm{Hz}$ and exhibited modulation spacing of approximately $8.5~\mathrm{Hz}$, close to the expected shaft-rotation frequency at $2.8~\mathrm{m\,s^{-1}}$ ($\approx 504~\mathrm{rpm}$). For Passage ID~3, a comparable component appeared near $966~\mathrm{Hz}$, with modulation spacing of approximately $5.5~\mathrm{Hz}$, consistent with the lower shaft speed at $1.8~\mathrm{m\,s^{-1}}$ ($\approx 324~\mathrm{rpm}$). The change in modulation spacing is therefore consistent with the rotational state of the propulsion system. However, because the center frequency did not scale linearly with shaft speed, the $\sim\!1~\mathrm{kHz}$ component is interpreted as a speed-sensitive tonal feature whose modulation is shaft-rate related, rather than as a simple shaft-order tone. A plausible interpretation is that this component reflects a mechanically or electromechanically mediated resonance whose excitation is influenced by propeller loading or shaft rotation, rather than a tone generated directly at the shaft frequency. Applying the Step~6 confidence criteria, this component is assigned tentative confidence as it recurs across passages, and its modulation spacing scales with shaft speed, but its center frequency does not scale linearly with shaft speed, leaving the physical mechanism incompletely identified.

A second prominent tonal group was observed near $5.5~\mathrm{kHz}$ (Fig.~\ref{fig:passby_spectrogram_5500}). The central component at approximately $5556~\mathrm{Hz}$ remained stable in both passages, consistent with a frequency imposed by the motor-drive control system rather than by shaft rotation. This frequency closely matches the velocity-loop update rate listed in Table~\ref{tab:auv_drive_specs}. Around this carrier, a dual sideband structure was observed. In Passage ID~2, the stronger sideband spacing was approximately $202~\mathrm{Hz}$, with a weaker spacing of approximately $101~\mathrm{Hz}$. In Passage ID~3, these spacings decreased to approximately $130~\mathrm{Hz}$ and $65~\mathrm{Hz}$, respectively. This speed dependence indicates motor-order modulation of a fixed carrier. The stronger spacing is consistent with the pole-passing rate of the 24-pole motor computed from~\eqref{eq:sideband_positions},
\begin{equation}
f_{\mathrm{pole}} = 24\,f_r,
\label{eq:pole_rate}
\end{equation}
where $f_r$ is the shaft-rotation frequency. For Passage ID~2,~\eqref{eq:pole_rate} gives $f_{\mathrm{pole}} = 201.6~\mathrm{Hz}$, close to the observed $\sim\!202~\mathrm{Hz}$ spacing; for Passage ID~3, $f_{\mathrm{pole}} = 129.6~\mathrm{Hz}$, in agreement with the observed $\sim\!130~\mathrm{Hz}$. The weaker spacing is consistent with the pole-pair rate of the same motor. Because a 24-pole motor contains 12 electrical pole pairs, the pole-pair rate --- which corresponds to the fundamental electrical frequency $f_e$ and may appear as a weaker sideband spacing due to drive or winding asymmetries is
\begin{equation}
f_{\mathrm{pair}} = 12\,f_r = \frac{1}{2}\,f_{\mathrm{pole}} = f_e.
\label{eq:pole_pair_rate}
\end{equation}
For Passage ID~2,~\eqref{eq:pole_pair_rate} gives $f_{\mathrm{pair}} = 100.8~\mathrm{Hz}$; for Passage ID~3, $f_{\mathrm{pair}} = 64.8~\mathrm{Hz}$, closely matching the observed weaker spacings of $101~\mathrm{Hz}$ and $65~\mathrm{Hz}$ respectively. The frequency residuals computed from~\eqref{eq:freq_residual} were within the spectral resolution for both passages, the component recurred across multiple passages, and the sideband spacings scaled with logged vehicle speed consistent with Eqs.~\eqref{eq:pole_rate} and~\eqref{eq:pole_pair_rate}. The $5.5~\mathrm{kHz}$ group is therefore assigned high confidence and interpreted as a fixed motor-drive-related carrier with speed-dependent motor-order modulation governed by both the pole-passing and pole-pair rates.

The same sideband structure recurred in the $11~\mathrm{kHz}$ region (Fig.~\ref{fig:passby_spectrogram_11000}). A stable component was observed near $11112~\mathrm{Hz}$, corresponding to the current-loop update rate and to the second harmonic of the $5556~\mathrm{Hz}$ component. The dominant and weaker sideband spacings decreased with speed in the same manner as in the $5.5~\mathrm{kHz}$ group, consistent with the pole-passing and pole-pair-rate interpretation. Frequency residuals were within the spectral resolution and the component recurred across passages; this group is therefore also assigned high confidence.

A stable component near $22224~\mathrm{Hz}$ was observed in the $22~\mathrm{kHz}$ region (Fig.~\ref{fig:passby_spectrogram_22000}), consistent with the PWM switching frequency and the fourth harmonic of $5556~\mathrm{Hz}$. This group exhibited the same speed-dependent sideband structure as the $5.5$ and $11~\mathrm{kHz}$ groups. The recurrence of comparable modulation around the $5.5$, $11$, and $22~\mathrm{kHz}$ carriers supports the interpretation that a common motor-order process modulates multiple controller-related frequencies. Frequency residuals were within the spectral resolution, and the component recurred consistently; this group is assigned high confidence.

\begin{figure}[!htbp]
    \centering
    \includegraphics[width=0.9\textwidth]{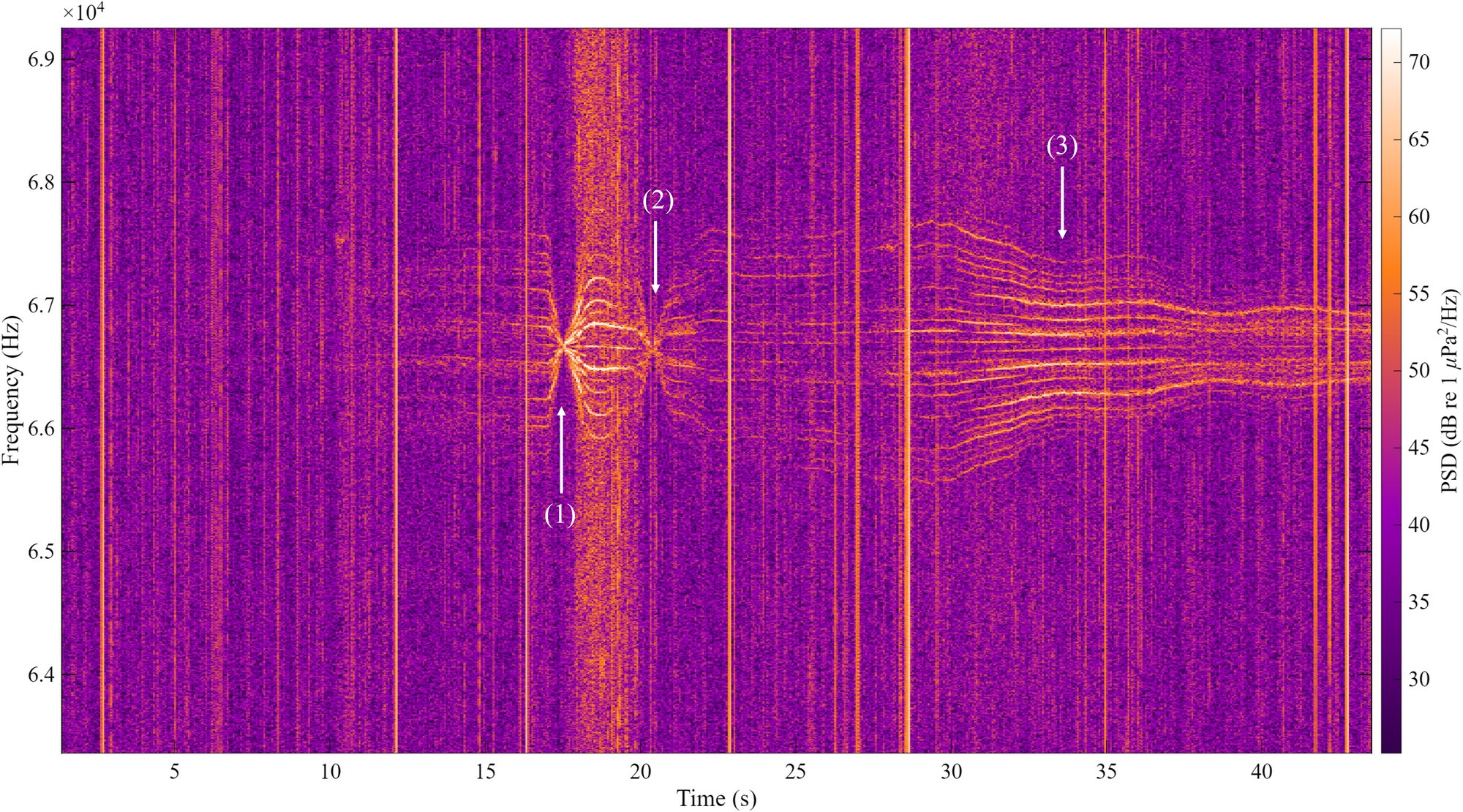}
    \caption{Spectrogram of a segment with rapid RPM changes prior to Passage ID~12 (256k FFT, Hann window, 50\% overlap, $63360$--$69250~\mathrm{Hz}$). The sequence includes a decrease from approximately $380~\mathrm{rpm}$ to 0~(1), an increase to approximately $486~\mathrm{rpm}$ and decrease to 0~(2), an increase to approximately $418~\mathrm{rpm}$, and a gradual decrease to approximately $270~\mathrm{rpm}$~(3).}
    \label{fig:passby_spectrogram_speedchange}
\end{figure}

Additional harmonic repetitions of the $5556$, $11112$, and $22224~\mathrm{Hz}$ groups were observed at higher frequencies, indicating a hierarchy of electronically imposed carrier frequencies and harmonics shaped by motor-order modulation. To further examine the link between sideband spacing and motor speed, a segment with rapid RPM changes prior to Passage ID~12 was analyzed (Fig.~\ref{fig:passby_spectrogram_speedchange}). The observed evolution of the sideband spacing during successive RPM changes was consistent with~\eqref{eq:sideband_evolution}, with the weaker pole-pair-rate component also present, providing additional support for the high-confidence attribution of the drive-related tonal groups.

\subsubsection{Step 7: Source-Related Tonal PSD Estimation}
\label{subsubsec:source_related_tonal_psd}

Applying Step~7, a PSD analysis was conducted for Passage IDs~2 and~3 to compare the received AUV spectrum with the local ambient reference and to identify tonal components suitable for propagation-corrected source-related estimation. Only components assigned high-confidence or tentative attribution in Step~6 and exceeding the ambient reference by at least $5~\mathrm{dB}$ were back-propagated. The PSD analysis used $1~\mathrm{s}$ Hann-windowed segments with $50\%$ overlap, giving an FFT-bin spacing of $1~\mathrm{Hz}$. The analysis interval was restricted to $\pm 15^{\circ}$ about CPA, rather than the broader $\pm 30^{\circ}$ sector commonly used in surface-vessel noise measurements~\cite{iso17208_1_2016,iso17208_2_2019}. This narrower interval was selected to emphasize the highest vehicle-to-ambient separation while limiting variation in range, observation aspect, and propagation conditions. For the close passages with $R_{\mathrm{CPA}} \approx 10~\mathrm{m}$, the maximum slant-range variation within this interval was approximately $0.35~\mathrm{m}$, which is smaller than the geometric uncertainty evaluated below.

\begin{figure}[!htbp]
    \centering
    \begin{subfigure}{0.9\textwidth}
        \centering
        \includegraphics[width=0.9\textwidth]{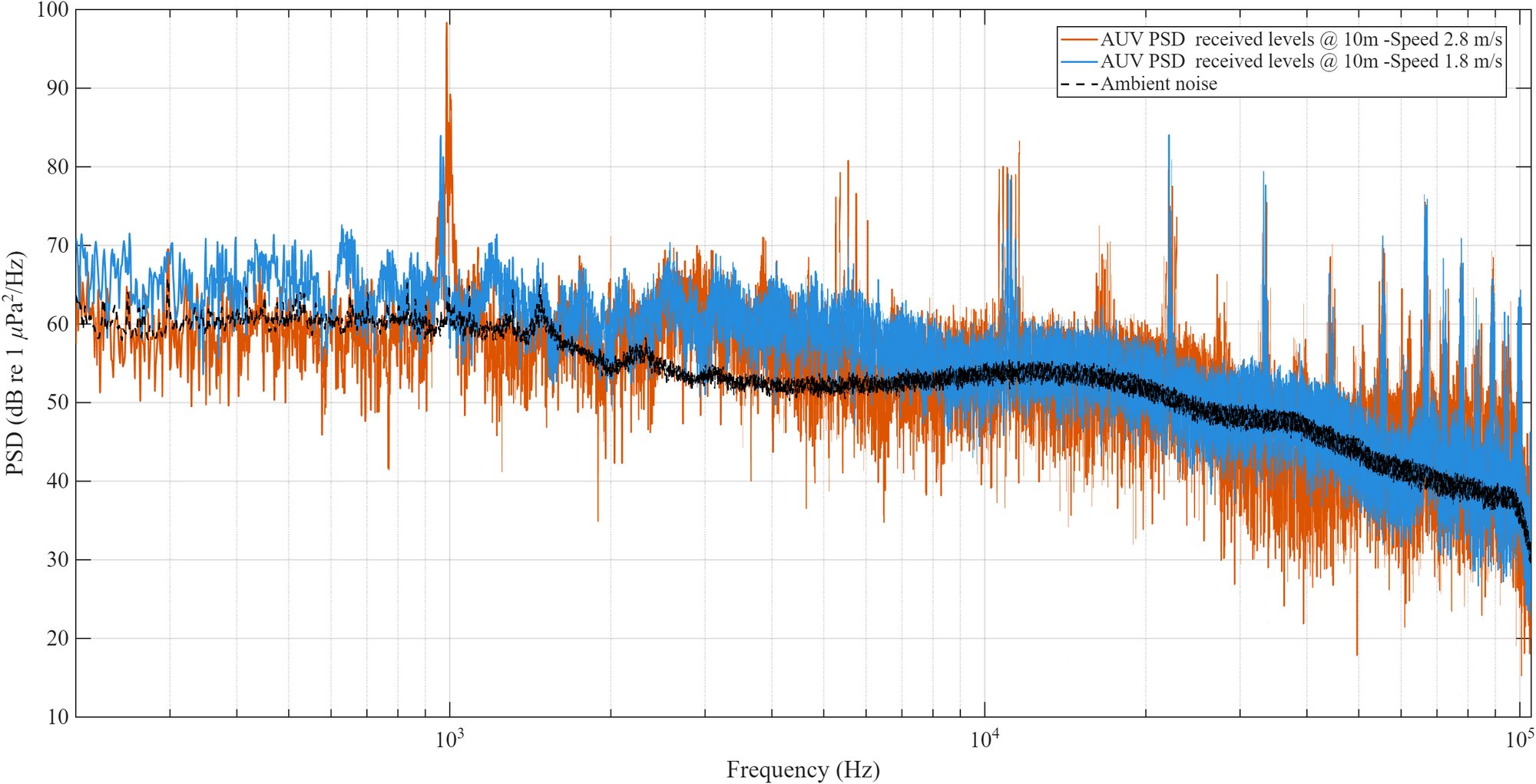}
        \caption{Received PSD with logarithmic frequency axis. The orange curve corresponds to Passage ID~2, the blue curve to Passage ID~3, and the black dashed line to the ambient reference.}
        \label{fig:psd_log_0832}
    \end{subfigure}
    \vspace{0.1cm}
    \begin{subfigure}{0.9\textwidth}
        \centering
        \includegraphics[width=0.9\textwidth]{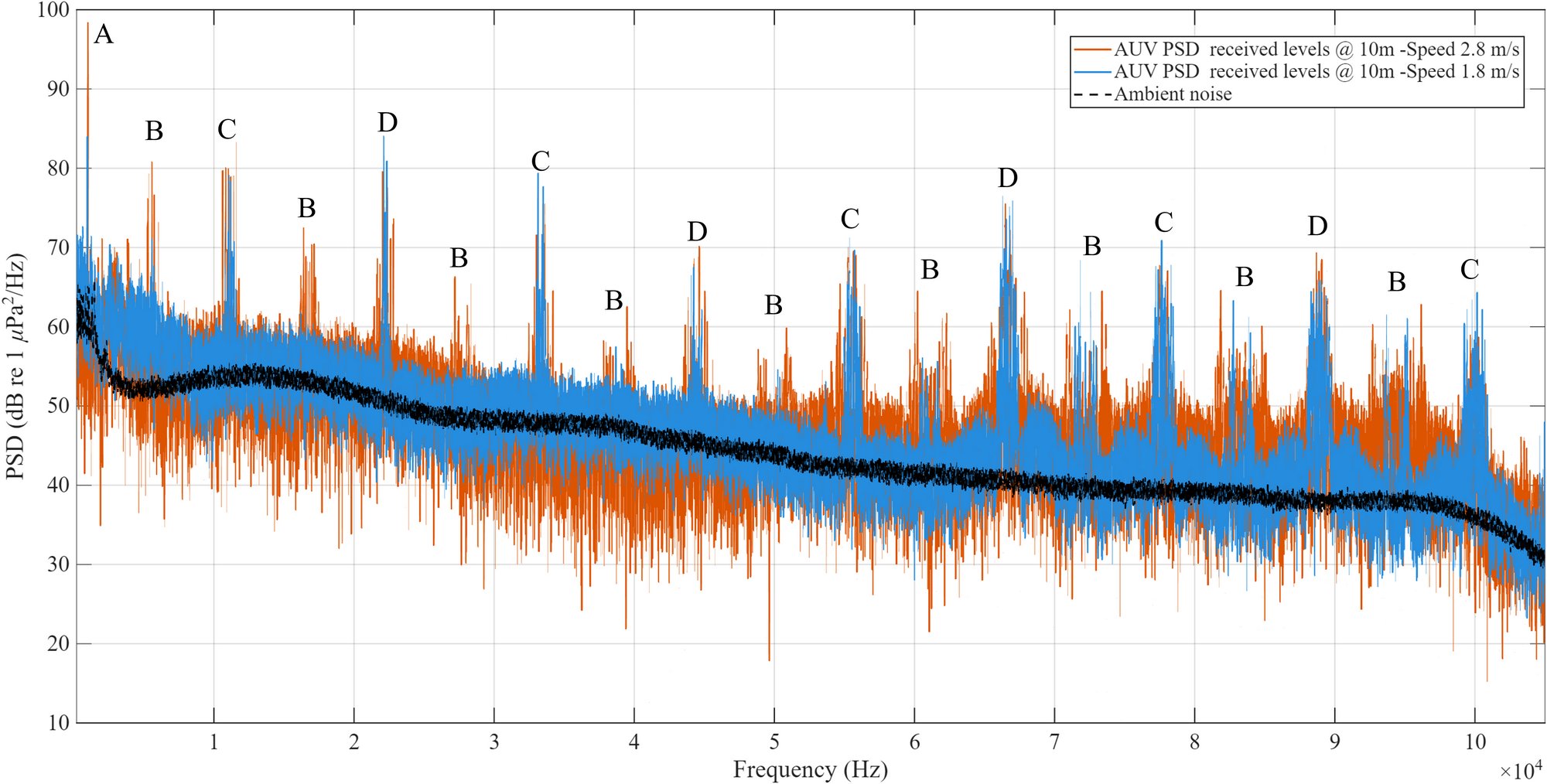}
        \caption{Received PSD with linear frequency axis. Tonal groups are labeled according to their Step~6 interpretation: A~---~$\sim\!1~\mathrm{kHz}$ tonal, B~---~$5.5~\mathrm{kHz}$-related series, C~---~$11~\mathrm{kHz}$-related series, D~---~$22~\mathrm{kHz}$ PWM-related series.}
        \label{fig:psd_lin_0832}
    \end{subfigure}
    \caption{Received PSD shown using (a)~a logarithmic frequency axis and (b)~a linear frequency axis. Orange: Passage ID~2; blue: Passage ID~3; black dashed: ambient reference.}
    \label{fig:psd_axes_comparison}
\end{figure}

The passage-specific ambient reference described in Section~\ref{subsubsec:step4_application} was used as the local baseline for the spectral comparisons. The received PSD spectra for Passage IDs~2 and~3 at CPA, together with the corresponding ambient reference, are shown in Fig.~\ref{fig:psd_axes_comparison}. The logarithmic-frequency representation (Fig.~\ref{fig:psd_log_0832}) facilitates inspection across the measured band, whereas the linear-frequency representation (Fig.~\ref{fig:psd_lin_0832}) highlights the discrete tonal components. In the linear plot, tonal groups are labeled according to their Step~6 interpretation: A denotes the $\sim\!1~\mathrm{kHz}$ tonal component, B the $5.5~\mathrm{kHz}$-related series, C the $11~\mathrm{kHz}$-related series, and D the $22~\mathrm{kHz}$ PWM-related series.

\begin{figure}[!htbp]
\centering
\includegraphics[width=0.9\textwidth]{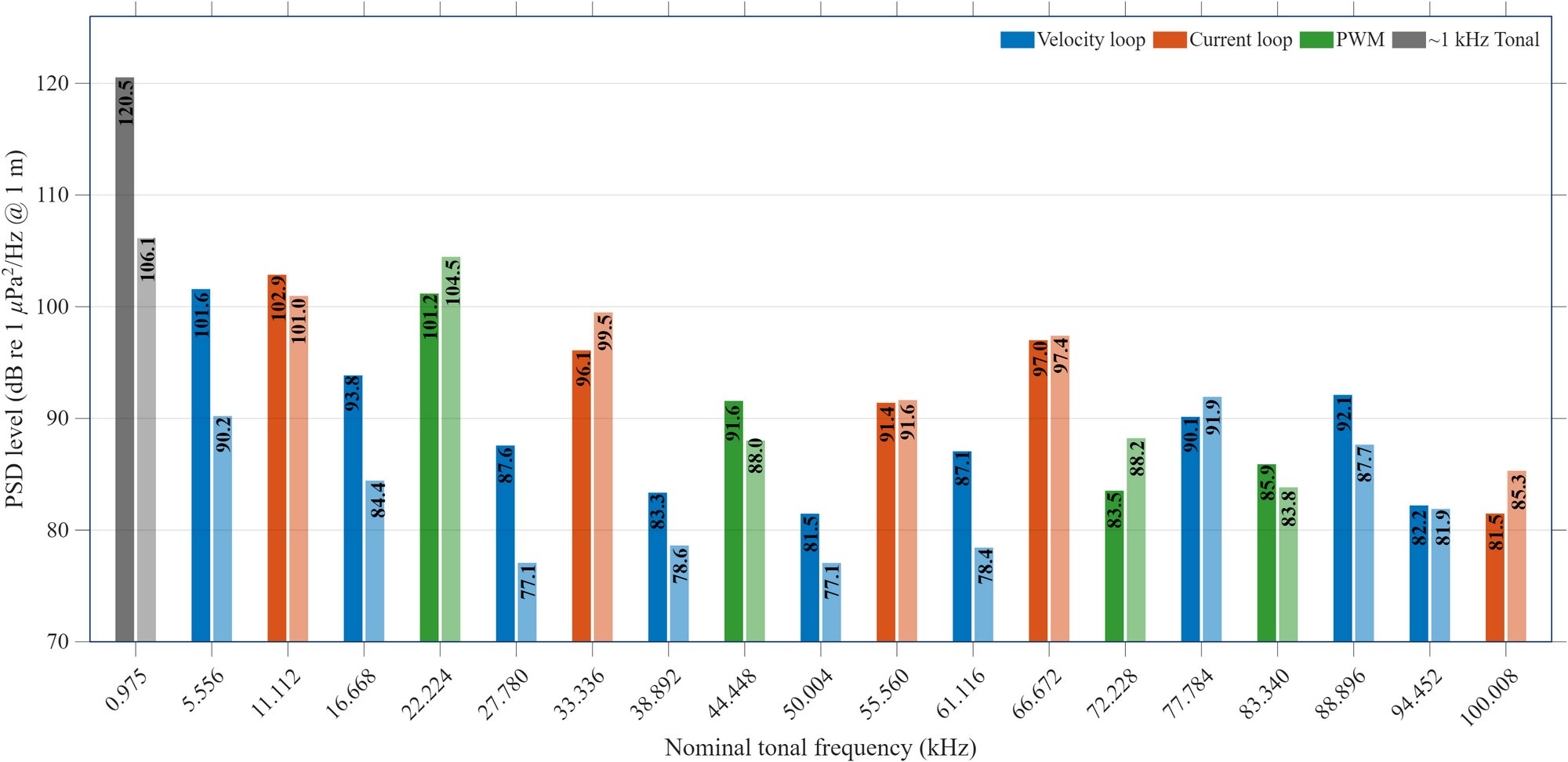}
\caption{Estimated source-related tonal PSD levels for Passage IDs~2 and~3 after band-dependent TL correction. Darker shades: $2.8~\mathrm{m\,s^{-1}}$; lighter shades: $1.8~\mathrm{m\,s^{-1}}$.}
\label{fig:sl_psd_compare}
\end{figure}

The received PSD exceeded the ambient reference only over limited parts of the measured band, primarily at discrete tonal frequencies and in selected broader regions. Elevated energy was evident in the $2$--$5~\mathrm{kHz}$ range and above approximately $50~\mathrm{kHz}$, whereas much of the spectrum below $50~\mathrm{kHz}$, outside the dominant tonal components, was only marginally above the ambient reference. Consistent with the Step~7 requirement, source-related estimation was restricted to identifiable tonal components that exceeded the ambient reference by at least $5~\mathrm{dB}$; ambient-limited regions were excluded from back-propagation. Transmission-loss corrections were computed using a close-range image-source propagation model configured to match the survey environmental parameters; the model specification and resulting TL field are described in~\ref{app:image_source}. Selected tonal peaks were back-propagated using~\eqref{eq:sl_psd_backprop_method}, where $L_{\mathrm{S,PSD}}$ is the source-related tonal PSD estimate in dB re $1~\mu\mathrm{Pa}^{2}/\mathrm{Hz}$ @ $1~\mathrm{m}$, $L_{\mathrm{R,PSD}}$ is the received PSD level in dB re $1~\mu\mathrm{Pa}^{2}/\mathrm{Hz}$, and $\mathrm{TL}(f)$ is the band-dependent transmission-loss correction. The resulting source-related tonal PSD estimates for both passages are shown in Fig.~\ref{fig:sl_psd_compare}.

The clearest speed-dependent change was observed for the tonal group near $1~\mathrm{kHz}$. The dominant component shifted from $997~\mathrm{Hz}$ in Passage ID~2 to $966~\mathrm{Hz}$ in Passage ID~3, while the source-related tonal PSD estimate decreased from approximately $120.5$ to $106.1~\mathrm{dB}$ re $1~\mu\mathrm{Pa}^{2}/\mathrm{Hz}$ @ $1~\mathrm{m}$, a reduction of approximately $14.4~\mathrm{dB}$. The modulation spacing associated with this component decreased from approximately $8.5$ to $5.5~\mathrm{Hz}$, consistent with the reduction in shaft-rotation rate. A similarly clear speed dependence was observed for the $5.56~\mathrm{kHz}$ velocity-loop-related tonal series. Its fundamental and identifiable harmonics exhibited reductions of approximately $8$--$11~\mathrm{dB}$ in the slower passage. Around the controller-related carriers, the sideband spacings decreased from approximately $202$ and $101~\mathrm{Hz}$ in Passage ID~2 to approximately $130$ and $65~\mathrm{Hz}$ in Passage ID~3, consistent with the pole-passing and pole-pair rates of the 24-pole motor established in Step~6. A plausible explanation for the reduced $5.56~\mathrm{kHz}$ levels is that the speed-control loop was less strongly excited as propeller loading decreased, reducing the amplitude of loop-related tonal content and its harmonics. By contrast, the $11.1~\mathrm{kHz}$ current-loop-related series and the $22~\mathrm{kHz}$ PWM-related series remained broadly stable between passages, with only modest and inconsistent variations. Applying the robustness criterion of~\eqref{eq:speed_dependence_criterion}, only level differences exceeding approximately $5~\mathrm{dB}$ were treated as robust, given the sensitivity of narrowband peaks to local spectral rippling, residual Doppler-related frequency shifts, and propagation-model uncertainty. Overall, the source-related PSD comparison indicates that the reduction in speed primarily affected the $\sim\!1~\mathrm{kHz}$ tonal and the $5.56~\mathrm{kHz}$ velocity-loop-related series, whereas the current-loop- and PWM-related tonal groups were comparatively insensitive to vehicle speed.

The geometric uncertainty was evaluated by applying Eqs.~\eqref{eq:range_uncertainty} and~\eqref{eq:geometric_uncertainty}. Evaluating the image-source TL model described in Appendix~\ref{app:image_source} at CPA ranges of $9$, $10$, and $11~\mathrm{m}$, and AUV depths of $19.5$ and $20.5~\mathrm{m}$, showed that a $\pm 1~\mathrm{m}$ variation in CPA range produced band-dependent TL deviations of up to approximately $\pm 1.5~\mathrm{dB}$. A $\pm 0.5~\mathrm{m}$ variation in source depth produced deviations generally within $\pm 0.5~\mathrm{dB}$. The combined geometric uncertainty $\delta L_S$, estimated as a root-sum-square from~\eqref{eq:geometric_uncertainty}, was therefore on the order of $\pm 1.5~\mathrm{dB}$. Additionally, as detailed in~\ref{app:supp_freq_band}, the Rayleigh distance for a $0.4~\mathrm{m}$ source reaches $21.9~\mathrm{m}$ at $105~\mathrm{kHz}$, exceeding the $10~\mathrm{m}$ CPA range; source-related estimates near the upper analysis limit should therefore be interpreted with additional caution. The Doppler analysis presented in~\ref{app:doppler} further indicated that the observed frequency residual was more consistent with a slightly higher AUV speed than with a reduced CPA range; the nominal logged range of approximately $10~\mathrm{m}$ was therefore retained, and the residual Doppler deviation is noted as a contributing factor to the overall geometric uncertainty.

\subsubsection{Step 8: Angular and Operational Analysis}
\label{subsubsec:step8_application}

Applying Step~8, angular and operational variability was assessed using two complementary comparisons: a receiver-geometry comparison between HYD1 and HYD2 at the same operating speed, and a before--after CPA comparison within a single close passage. In both cases, the robustness criteria of Eqs.~\eqref{eq:aspect_difference_criterion} and~\eqref{eq:speed_dependence_criterion} were applied using the combined geometric uncertainty $\delta L_S \approx \pm 1.5~\mathrm{dB}$ established in Step~7.

Source-related tonal PSD estimates were compared between the two receiver configurations for Passage IDs~1 and~2, both recorded at $2.8~\mathrm{m\,s^{-1}}$. HYD1 provided a near-under-track geometry at CPA, while HYD2 was laterally offset from the track. Received tonal PSD levels were corrected using the same band-dependent TL procedure described in Section~\ref{subsubsec:source_related_tonal_psd}, and source-related estimates were calculated using~\eqref{eq:sl_psd_backprop_method}. The purpose of the comparison was to evaluate whether the dominant tonal groups identified in Step~6 were repeatable across receiver configurations, rather than to estimate a complete directivity pattern.

\begin{figure}[!htbp]
\centering
\includegraphics[width=0.9\textwidth]{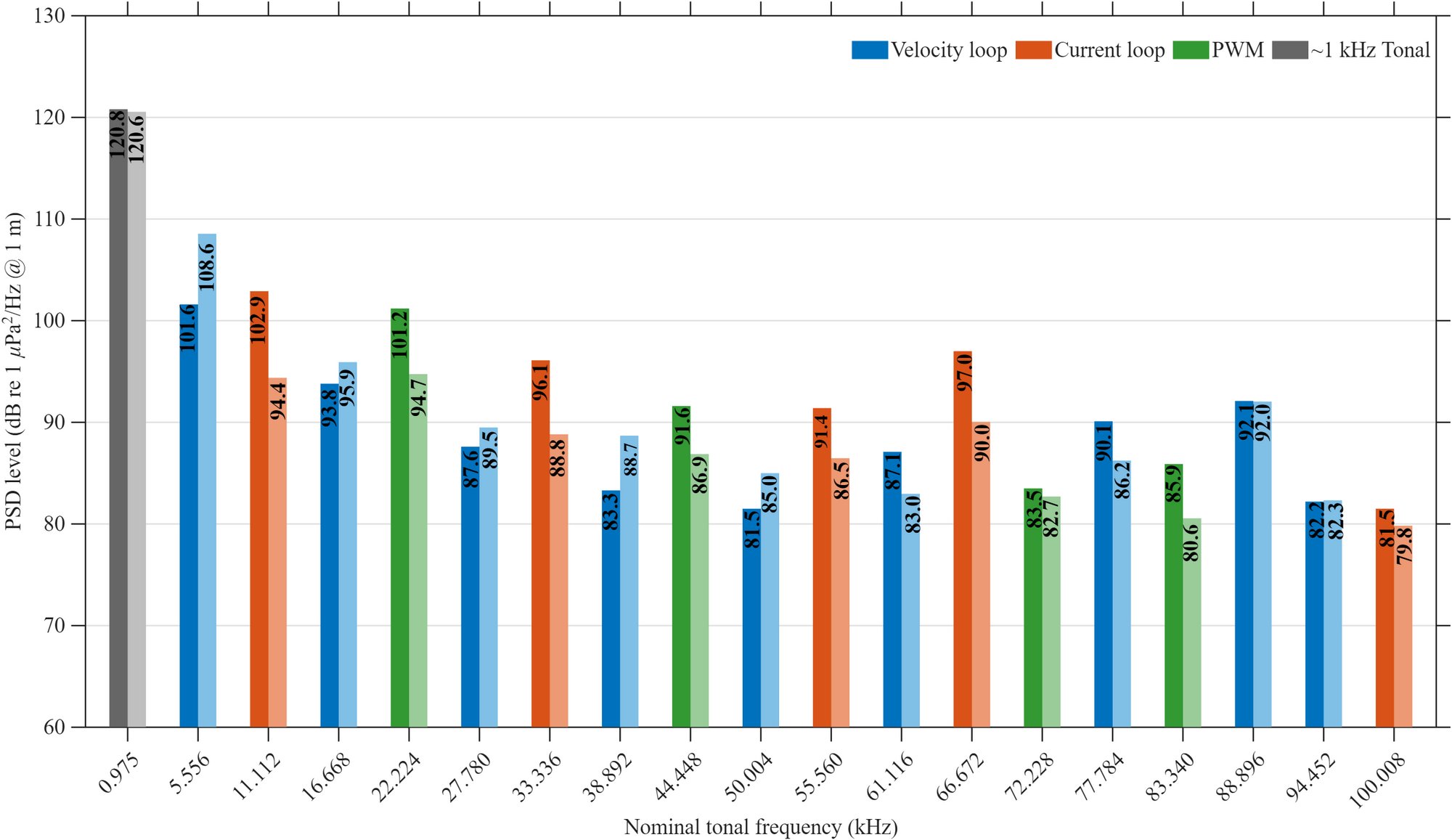}
\caption{Estimated source-related tonal PSD levels from HYD1 and HYD2 after band-dependent TL correction. HYD1 represents a near-under-track receiver configuration, while HYD2 represents a laterally offset configuration. Both passages were recorded at $2.8~\mathrm{m\,s^{-1}}$.}
\label{fig:sl_psd_compare_hyd}
\end{figure}

The main tonal groups were present in both receiver configurations (Fig.~\ref{fig:sl_psd_compare_hyd}), confirming that the dominant spectral components identified in Step~6 were not specific to a single receiver location. To apply the robustness criterion of~\eqref{eq:aspect_difference_criterion}, a range-specific geometric uncertainty was evaluated separately for each receiver. For HYD1 at $r \approx 10~\mathrm{m}$, the combined uncertainty $\delta L_{S,1}$ was established in Step~7 as approximately $\pm 1.5~\mathrm{dB}$ (band-maximum $\pm 2.0~\mathrm{dB}$). For HYD2 at $r \approx 78~\mathrm{m}$, the image-source TL model was recomputed at ranges of $77.4$ and $79.4~\mathrm{m}$ and at source depths of $19.5$ and $20.5~\mathrm{m}$; the resulting band-dependent combined uncertainty $\delta L_{S,2}$ had a mean of approximately $\pm 1.5~\mathrm{dB}$ and a maximum of approximately $\pm 2.7~\mathrm{dB}$ across the analysis band. The combined threshold of~\eqref{eq:aspect_difference_criterion} is therefore
\begin{equation}
    \sqrt{\delta L_{S,1}^{2}+\delta L_{S,2}^{2}}
    \approx 2.1~\mathrm{dB}\ \text{to}\ 3.4~\mathrm{dB},
    \label{eq:hyd_combined_threshold}
\end{equation}
where $2.1~\mathrm{dB}$ and $3.4~\mathrm{dB}$ correspond to band-mean and band-maximum values respectively.
Applying this band-dependent threshold, no individual tonal difference between HYD1 and HYD2 consistently exceeded the criterion across the analysis band. The HYD1/HYD2 comparison is therefore treated as confirmation of tonal repeatability across receiver locations. Aspect-related variability is assessed only from the before--after CPA comparison in Passage~ID~2, where the source--receiver range is controlled within a single receiver.

To examine apparent aspect-related variability within a single close passage, received levels were compared at symmetric time offsets before and after CPA for Passage ID~2. Because the AUV speed was approximately constant, equal time offsets correspond to approximately equal source--receiver ranges, satisfying the Step~8 requirement of~\eqref{eq:aspect_window}. Residual contributions from multipath propagation, Doppler shift, source-depth variation, and local interference may also affect the observed differences; accordingly, results are reported as before--after received-level asymmetries rather than as source directivity estimates.

Spectrograms were computed using $0.5~\mathrm{s}$ analysis windows with a $0.25~\mathrm{s}$ step size and an FFT length of 524288 samples. Band-integrated received levels were computed using $0.25~\mathrm{s}$ windows with a $0.125~\mathrm{s}$ step size and an FFT length of 262144 samples. These settings provided a dense frequency grid for tracking narrowband tonal components near CPA, although the effective frequency resolution is primarily determined by the finite window duration rather than the zero-padded FFT length.

\begin{figure}[!htbp]
\centering
\includegraphics[width=0.9\textwidth]{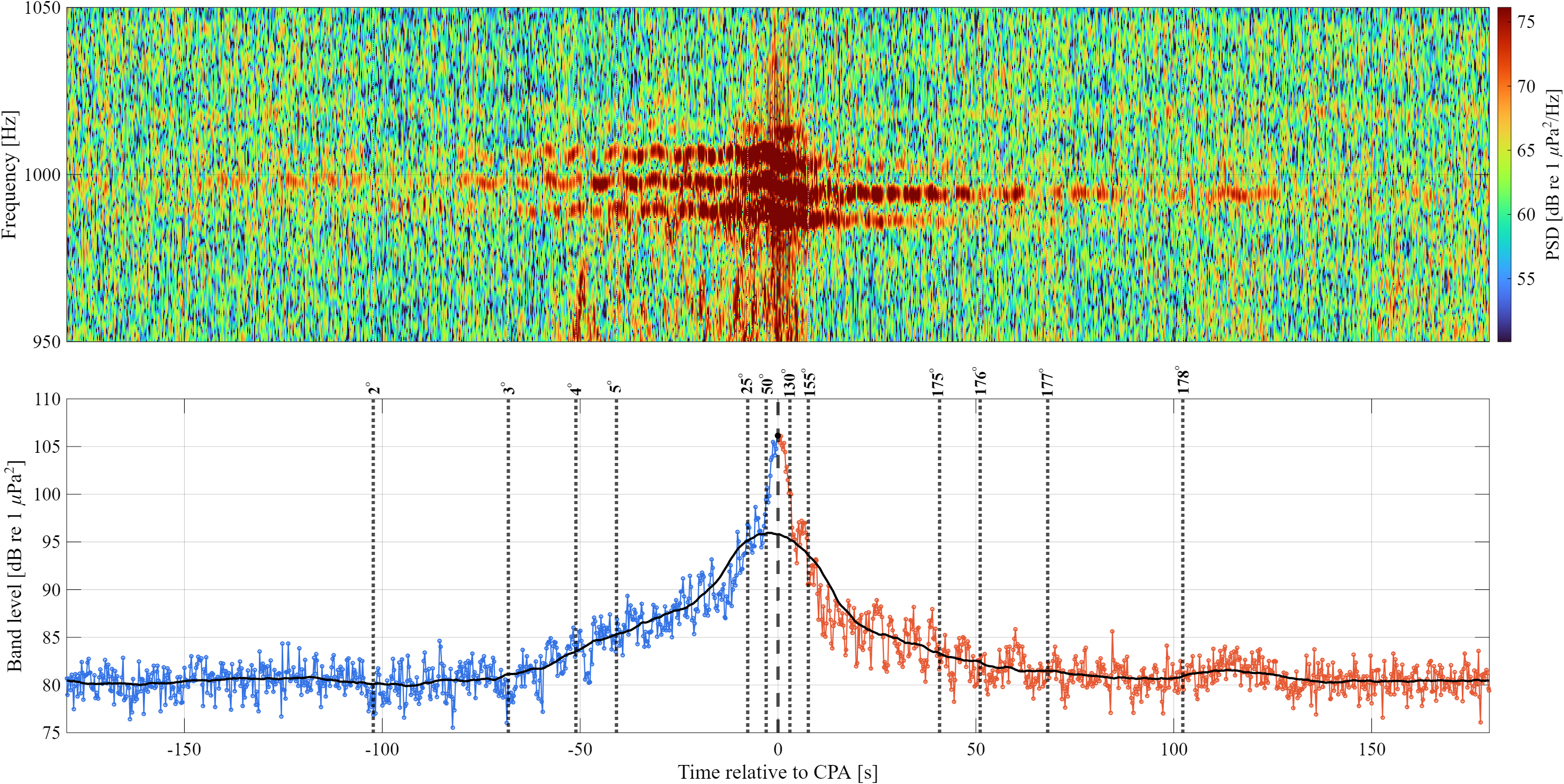}
\caption{Spectrogram and band-integrated received level around the $1~\mathrm{kHz}$ tonal component. Upper panel: PSD spectrogram. Lower panel: received level integrated over $970$--$1020~\mathrm{Hz}$. Time is relative to CPA. Blue and red markers represent windows before and after CPA, respectively; the black curve shows the smoothed trend. Spectrogram: $0.5~\mathrm{s}$ window, $0.25~\mathrm{s}$ step, 524288-point FFT. Received level: $0.25~\mathrm{s}$ window, $0.125~\mathrm{s}$ step, 262144-point FFT.}
\label{fig:angular_dependence_1000hz}
\end{figure}

Figure~\ref{fig:angular_dependence_1000hz} shows the spectrogram and band-integrated received level for the $970$--$1020~\mathrm{Hz}$ band. Mean received levels before and after CPA were $83.0$ and $82.7~\mathrm{dB}$ respectively, corresponding to a mean paired difference of $-0.34~\mathrm{dB}$ and an energy-averaged difference of approximately $-0.05~\mathrm{dB}$. This difference is well below the robustness threshold of $2\,\delta L_S \approx 3~\mathrm{dB}$ from~\eqref{eq:speed_dependence_criterion}, and the $970$--$1020~\mathrm{Hz}$ component is therefore classified as inconclusive with respect to aspect-related variability during this passage.

\begin{figure}[!htbp]
\centering
\includegraphics[width=0.9\textwidth]{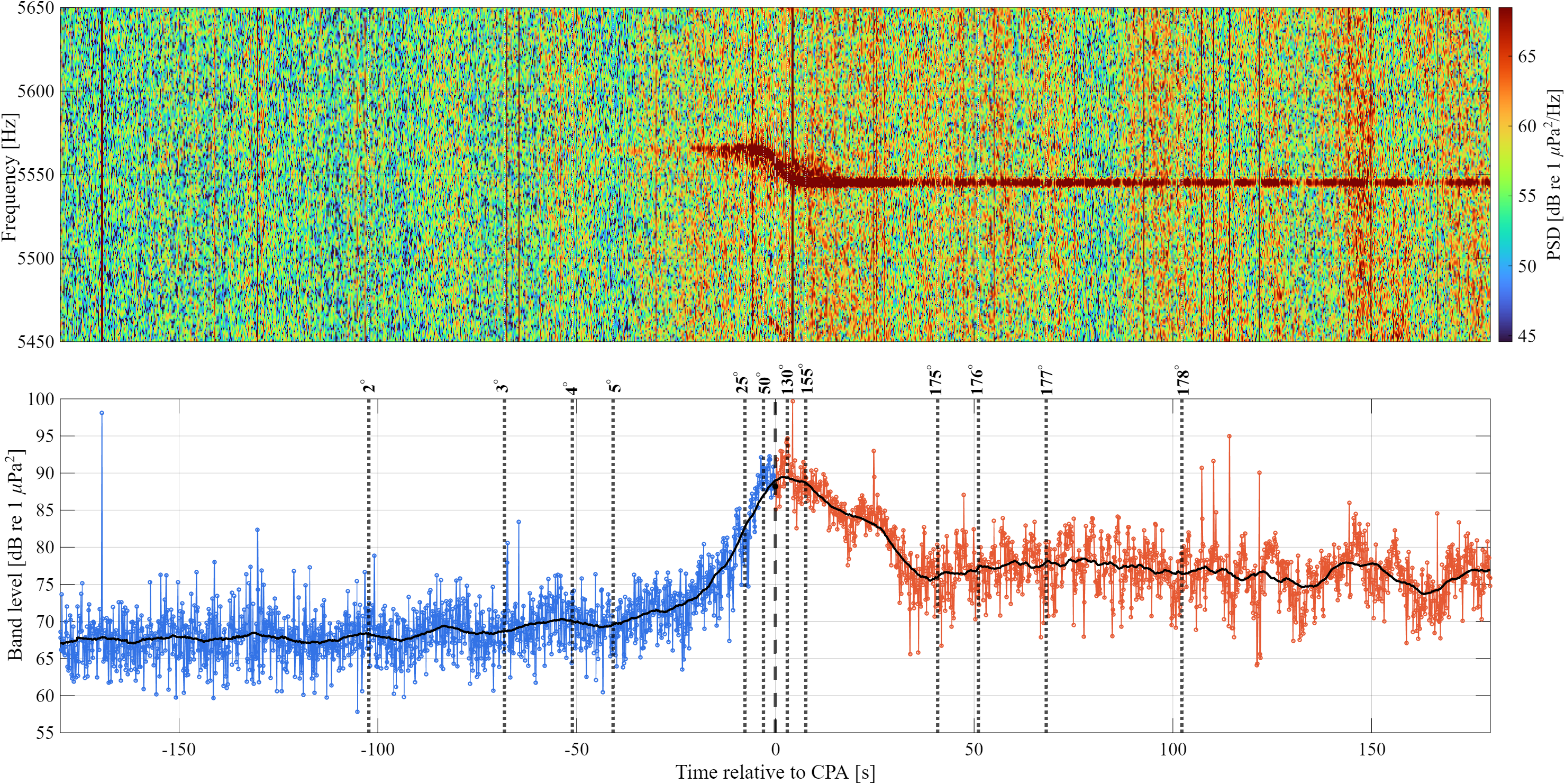}
\caption{Spectrogram and band-integrated received level around the $5.556~\mathrm{kHz}$ tonal component. Upper panel: PSD spectrogram. Lower panel: received level integrated over the Doppler-adjusted tonal band. Time is relative to CPA. Blue and red markers represent windows before and after CPA respectively; black curve shows the smoothed trend. Spectrogram: $0.5~\mathrm{s}$ window, $0.25~\mathrm{s}$ step, 524288-point FFT. Received level: $0.25~\mathrm{s}$ window, $0.125~\mathrm{s}$ step, 262144-point FFT.}
\label{fig:angular_dependence_5556hz}
\end{figure}

Figure~\ref{fig:angular_dependence_5556hz} shows the time evolution of the $5.556~\mathrm{kHz}$ tonal component. Because the tonal frequency varied slightly during the passage due to Doppler shift, the received level was integrated over $5550$--$5570~\mathrm{Hz}$ before CPA and $5540$--$5560~\mathrm{Hz}$ after CPA. The mean received band level increased from $69.9~\mathrm{dB}$ re $1~\mu\mathrm{Pa}^{2}$ before CPA to $78.2~\mathrm{dB}$ re $1~\mu\mathrm{Pa}^{2}$ after CPA, an average difference of approximately $8.3~\mathrm{dB}$. Mean PSD level similarly increased from $57.2$ to $65.2~\mathrm{dB}$ re $1~\mu\mathrm{Pa}^{2}/\mathrm{Hz}$. This difference exceeds the robustness threshold of $2\,\delta L_S \approx 3~\mathrm{dB}$ from~\eqref{eq:speed_dependence_criterion}, and the before--after asymmetry at $5.556~\mathrm{kHz}$ is therefore classified as robust, indicating an apparent aspect-related received-level increase on the post-CPA side.

\begin{figure}[!htbp]
\centering
\includegraphics[width=0.9\textwidth]{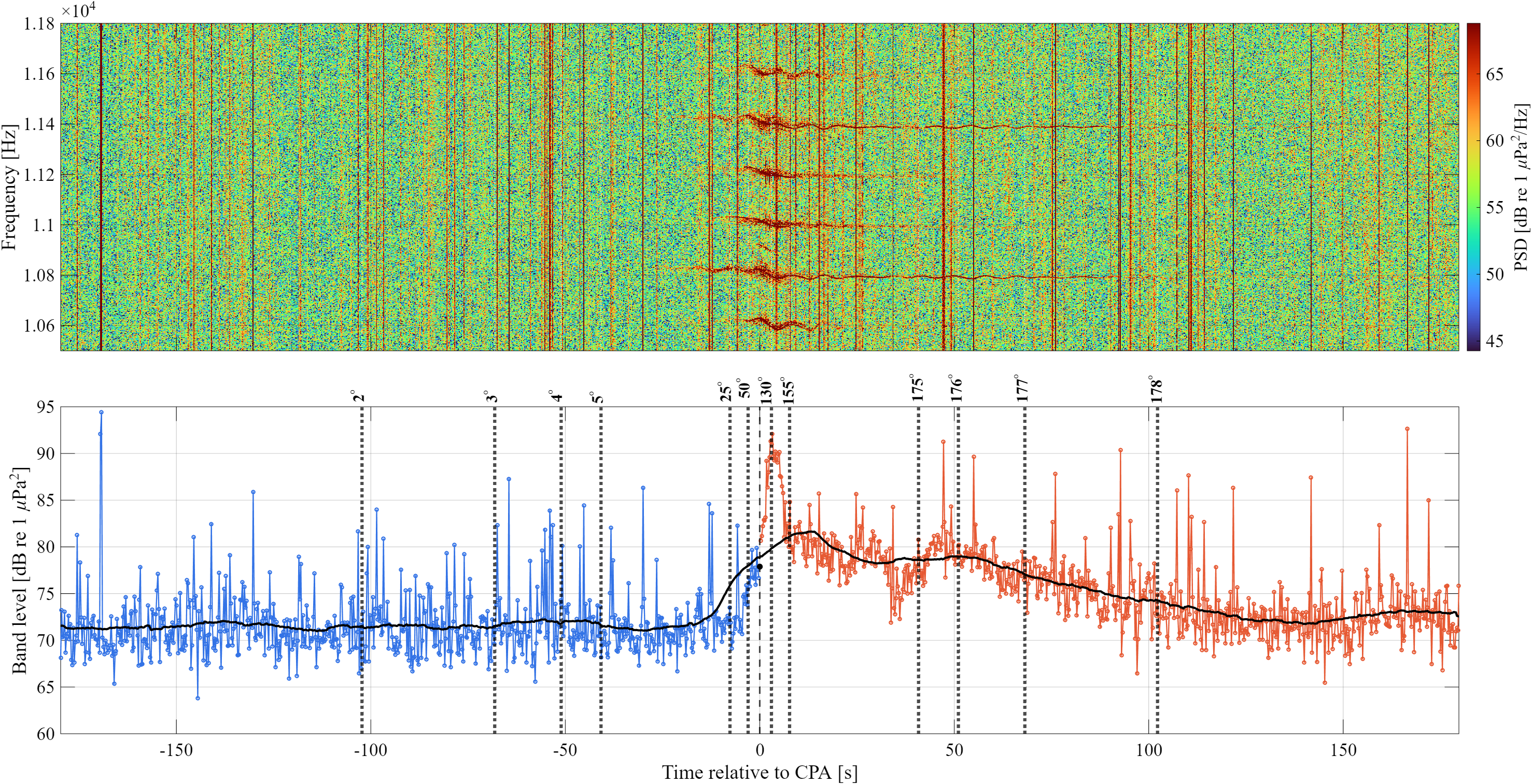}
\caption{Spectrogram and band-integrated received level around the $11~\mathrm{kHz}$ tonal region. Upper panel: PSD spectrogram. Lower panel: received level integrated over the selected tonal band. Time is relative to CPA. Blue and red markers represent windows before and after CPA respectively; black curve shows the smoothed trend. Spectrogram: $0.5~\mathrm{s}$ window, $0.25~\mathrm{s}$ step, 524288-point FFT. Received level: $0.25~\mathrm{s}$ window, $0.125~\mathrm{s}$ step, 262144-point FFT.}
\label{fig:angular_dependence_11khz}
\end{figure}

A similar, though smaller, asymmetry was observed in the $11~\mathrm{kHz}$ region (Fig.~\ref{fig:angular_dependence_11khz}), where tonal components near $10.8$ and $11.4~\mathrm{kHz}$ were generally higher after CPA. For the component around $10.8~\mathrm{kHz}$, the mean paired after--before difference was approximately $4.1~\mathrm{dB}$, and the energy-averaged difference was approximately $4.5~\mathrm{dB}$, with after-CPA levels higher in about $80\%$ of paired comparisons. The asymmetry was strongest near CPA, where after-CPA levels exceeded before-CPA levels by approximately $10.9~\mathrm{dB}$ within the first $10~\mathrm{s}$, and decreased gradually at larger time offsets. This difference exceeds the robustness threshold, and the before--after asymmetry at $10.8~\mathrm{kHz}$ is therefore also classified as robust.

The before--after comparison therefore indicates frequency-dependent aspect-related variability: the $970$--$1020~\mathrm{Hz}$ component was inconclusive, whereas the $5.556~\mathrm{kHz}$ and $10.8$--$11.4~\mathrm{kHz}$ components showed robust before--after asymmetry with higher received levels after CPA. Because the comparisons were made at equal time offsets corresponding to approximately equal source--receiver ranges, the observed differences are unlikely to result from range-dependent transmission loss alone. However, the available single-receiver geometry does not allow propagation effects and true source directivity to be fully separated, and the results are therefore reported as received-level asymmetries consistent with Step~8 requirements.

\section{Discussion}
\label{sec:discussion}

\subsection{Signature Interpretation and Speed Dependence}
\label{subsec:discussion_signature}

The Step~6 analysis identified two distinct categories of tonal component. The stable groups near $5556$, $11112$, and $22224~\mathrm{Hz}$ were assigned high confidence and attributed to velocity-loop, current-loop, and PWM-related carrier frequencies, with sideband spacings tracking the pole-passing and pole-pair rates of the 24-pole motor. Their recurrence across multiple carriers and independent passages confirms that motor-order modulation is a structural feature of the drive-related signature rather than a measurement artifact. The $\sim\!1~\mathrm{kHz}$ component was assigned tentative confidence: its modulation spacing followed the shaft-rotation rate, but the center frequency did not scale linearly with shaft speed, leaving the physical mechanism — possibly a structural or electromechanical resonance driven by propeller loading or drivetrain vibration — incompletely identified.

Speed reduction from $2.8$ to $1.8~\mathrm{m\,s^{-1}}$ produced selective rather than uniform changes. The $\sim\!1~\mathrm{kHz}$ tonal and the $5.56~\mathrm{kHz}$ velocity-loop series decreased by approximately $14.4$ and $8$--$11~\mathrm{dB}$ respectively, consistent with reduced speed-control excitation as propeller loading decreased. By contrast, the $11.1~\mathrm{kHz}$ and $22~\mathrm{kHz}$ groups remained broadly stable, with level differences below the $2\,\delta L_S$ threshold of~\eqref{eq:speed_dependence_criterion}. This selective behavior shows that broadband speed scaling is insufficient to characterize SEV radiated noise; different tonal families governed by different subsystems may respond differently to changes in operating state.

\subsection{Aspect-Related Variability and Receiver Geometry}
\label{subsec:discussion_aspect_geometry}
The receiver-geometry comparison confirmed that the dominant tonal groups were repeatable between the near-under-track (HYD1) and laterally offset (HYD2) configurations at $2.8~\mathrm{m\,s^{-1}}$, supporting their origin in the AUV rather than in receiver-specific propagation conditions. A range-specific uncertainty analysis showed that the combined geometric uncertainty for HYD2 (band-mean $\pm 1.5~\mathrm{dB}$, band-maximum $\pm 2.7~\mathrm{dB}$) is comparable to, and in several frequency bands exceeds, that of HYD1, yielding a combined robustness threshold of $2.1$--$3.4~\mathrm{dB}$ from~\eqref{eq:hyd_combined_threshold}. No observed spectral difference between the two receivers was robust under this band-dependent criterion. The HYD1/HYD2 comparison, therefore, supports tonal repeatability across receiver locations but cannot be used as evidence of aspect-dependent radiation; that assessment rests solely on the before-and-after CPA comparison reported below.

The before-and-after CPA comparison revealed frequency-dependent behavior. The $970$--$1020~\mathrm{Hz}$ component showed a mean asymmetry of $-0.34~\mathrm{dB}$, well below the robustness threshold and classified as inconclusive. The $5.556~\mathrm{kHz}$ and $10.8$--$11.4~\mathrm{kHz}$ components showed robust asymmetries of approximately $8.3$ and $4.1$--$4.5~\mathrm{dB}$ respectively, with after-CPA levels consistently higher. Because comparisons were made at equal time offsets corresponding to approximately equal source--receiver ranges, range-dependent spreading is an unlikely explanation, suggesting true aspect-dependent radiation. A natural physical interpretation of the after-CPA asymmetry is the stern-aspect geometry: because the AUV propulsion system (i.e., motor and propeller) is located at the stern, the post-CPA half of each transit corresponds to the hydrophone observing the propulsor end of the vehicle. Drive-related motor tones are expected to be stronger in this direction, since the motor and drive electronics are located at the stern and the propeller provides less acoustic shielding in the aft aspect than the vehicle hull does in the forward aspect. These findings are consistent with previous AUV directivity observations~\cite{gebbie2012aspect} and extend them to higher-frequency motor-drive components. The Step~8 uncertainty thresholds provided a principled basis for distinguishing robust from inconclusive asymmetries, which is an improvement over the qualitative comparisons in earlier studies.

\subsection{Implications, Limitations, and Future Work}
\label{subsec:discussion_implications}

The test case demonstrates that the framework provides a reproducible, traceable workflow for SEV URN characterization. Each decision — CPA geometry, frequency band, ambient reference, tonal attribution, propagation correction, and robustness classification — is linked to explicit equations and criteria, addressing the methodological fragmentation identified in Section~\ref{sec:existing_characterization}.

Two enabling requirements stand out. First, vehicle metadata was essential: the Step~6 attribution was possible only because the motor pole number, PWM frequency, controller rates, and the speed-to-RPM relationship were documented beforehand. Without this information, the observed tonal and sideband features would remain physically unattributed. Second, recording bandwidth must be matched to the expected drive frequencies. The principal tonal groups in this case fell near $5.56$, $11.1$, and $22.2~\mathrm{kHz}$ — well above the low-frequency range typically used in vessel-noise studies — motivating the systematic upper-bound selection of~\eqref{eq:fmax_pwm} before deployment.

Source-related estimates carry combined geometric uncertainties of approximately $\pm 1.5~\mathrm{dB}$, with additional contributions from seabed properties, surface conditions, and residual Doppler offsets. The estimates are therefore most useful for comparing tonal components between passages and operating states rather than for defining an absolute platform source level. The ten excluded transits also highlight the practical importance of close CPA ranges, passage-specific ambient references, and careful contamination screening when measuring low-level SEV signatures in coastal environments.

Future work should apply the framework to additional SEV platforms under controlled speed and depth conditions, using repeated reciprocal tracks and multi-receiver geometries with comparable slant ranges. This would enable stronger separation of source directivity, propagation effects, and operational variability and would support higher-confidence Step~8 assessments. The stability of motor-order modulation patterns across different payload configurations, propeller types, and controller settings also warrants further investigation. The framework currently relies on Welch-based spectral estimation, which is subject to the conventional resolution limit of $1/T_w$. For signals with well-defined tonal components, subspace-based super-resolution methods such as the matrix pencil~\cite{hua1990matrixpencil}, MUSIC~\cite{schmidt1986music}, or ESPRIT~\cite{roy1989esprit} could in principle resolve sideband spacings and track Doppler frequency trajectories from shorter analysis windows than the DFT requires, providing denser time resolution for monitoring speed-dependent modulation patterns. Their practical application to SEV URN data is, however, complicated by the need to specify the number of sinusoidal components in advance and by the non-stationary ambient-noise background typical of coastal environments.

\section{Conclusions}
\label{sec:conclusions}

This paper presented an eight-step methodology for characterizing URN from submerged electric vehicles in coastal waters and demonstrated it using an A18D AUV as a test case. The framework integrates calibrated pass-by measurements, synchronized vehicle navigation data, ambient noise assessment, propagation-corrected source-related tonal PSD estimation, subsystem-oriented spectral interpretation, and angular and operational assessment. Each step is defined by explicit input requirements, decision criteria, and documented outputs, providing a traceable and reproducible basis for SEV acoustic characterization that is not available from broadband or ad hoc measurement approaches.

The test case demonstrated the framework's practical applicability under realistic coastal field conditions. Applying Step~2, cavitation was screened and classified as unlikely at all operating speeds, permitting tonal and motor-drive-related interpretation to proceed. Applying Step~3, the analysis band of $200~\mathrm{Hz}$--$105~\mathrm{kHz}$ was derived from the drive metadata and survey geometry, capturing all expected PWM harmonics and motor-order sidebands. Applying Steps~5 and~6, the measured AUV signature was dominated by discrete tonal, harmonic, and modulated components. Three tonal families --- centered near $5.56$, $11.1$, and $22.2~\mathrm{kHz}$ --- were assigned high confidence and attributed to velocity-loop, current-loop, and PWM-related carrier frequencies, respectively, with speed-dependent sideband spacings governed by the pole-passing and pole-pair rates of the 24-pole motor. A fourth component near $1~\mathrm{kHz}$ was assigned tentative confidence, with shaft-rate modulation but an incompletely identified center-frequency mechanism. Applying Step~7, source-related tonal PSD estimates ranged from approximately $77$ to $120~\mathrm{dB}$ re $1~\mu\mathrm{Pa}^{2}/\mathrm{Hz}$ @ $1~\mathrm{m}$, with a combined geometric uncertainty of approximately $\pm 1.5~\mathrm{dB}$. Speed reduction from $2.8$ to $1.8~\mathrm{m\,s^{-1}}$ produced selective changes: the $\sim\!1~\mathrm{kHz}$ and $5.56~\mathrm{kHz}$ tonal families decreased substantially, while the $11.1$ and $22~\mathrm{kHz}$ groups remained broadly stable. Applying Step~8, before--after CPA comparisons classified the $5.556~\mathrm{kHz}$ and $10.8$--$11.4~\mathrm{kHz}$ asymmetries as robust, while the $\sim\!1~\mathrm{kHz}$ asymmetry was inconclusive.

The results show that effective SEV characterization requires narrowband analysis, detailed propulsion and control-system metadata, and cautious interpretation of source-related estimates within a structured uncertainty framework. They also show that PWM-driven SEVs may radiate important tonal components well above the low-to-mid-frequency range emphasized in previous studies. Future work should apply the framework to additional SEV platforms, under controlled operating conditions, with repeated passes, and with multi-receiver geometries to better quantify source variability, frequency-dependent directivity, and the generality of the motor-order modulation patterns identified here.

\section*{CRediT authorship contribution statement}

\noindent Mark Shipton: Methodology, Software, Formal analysis, Investigation, Data curation, Writing -- original draft, Visualization.

\noindent Amir Boag: Writing -- review \& editing, Funding acquisition, Conceptualization.

\noindent Roee Diamant: Conceptualization, Methodology, Writing -- review \& editing, Supervision, Project administration, Funding acquisition.

\section*{Declaration of competing interest}
The authors declare that they have no known competing financial interests or personal relationships that could have appeared to influence the work reported in this paper.

\section*{Funding}
This research was supported by a scholarship funded by the Israel Science Foundation [grant number 973/23], the Mediterranean Sea Research Center of Israel (MERCI) at the University of Haifa, and the European Union's Horizon Europe program under the UWIN-LABUST project [grant number 101086340].

\section*{Data Availability}
Data supporting this study, including the synchronized vehicle metadata and representative processed acoustic data sufficient to reproduce the figures and numerical results reported in this paper, are publicly available in the repository.

\appendix

\section{Selection of the Analysis Frequency Band for the AUV test case}
\label{app:supp_freq_band}

The lower frequency bound of \(200~\mathrm{Hz}\) adopted in the main analysis was motivated by two considerations: the acoustic near-field transition at the close CPA geometry, and the influence of shallow-water propagation effects on low-frequency source-related estimates. Both are described below.

\subsection{Near-Field Transition Criterion}

Source-related estimation assumes far-field propagation, in which the received level decays predictably with range. At distances shorter than approximately one acoustic wavelength, or shorter than three times the largest characteristic source dimension, this assumption may not hold, and back-propagated estimates become unreliable~\cite{foote2014nearfield_farfield,irs2025urn}. The near-field distance is approximated as
\begin{equation}
r_{\mathrm{NF}} \approx \max\!\left(\frac{c_w}{f},\; 3D,\; \frac{2D^2 f}{c_w}\right),
\label{eq:supp_near_field_distance}
\end{equation}
where \(\lambda = c_w / f\) is the acoustic wavelength and \(D\) is a characteristic source dimension. This expression is used here as a practical screening criterion rather than a fundamental acoustic boundary.

The effective radiating dimension \(D\) of a submerged vehicle is not limited to the motor alone. Vibration generated by the motor, shaft, propeller, and drivetrain may be transmitted through the hull and radiated from a larger area of the vehicle body~\cite{ross1987mechanics,jensen2011computational}. To bound this uncertainty, two representative values of \(D\) are considered: the motor axial dimension \(D = 0.20~\mathrm{m}\), and the propeller diameter \(D = 0.40~\mathrm{m}\). Table~\ref{tab:nearfield_sensitivity} summarizes the resulting near-field distances and implied lower frequency bounds for each assumed value of \(D\) at the \(10~\mathrm{m}\) CPA geometry.

\begin{table}[!htbp]
\centering
\footnotesize
\caption{Sensitivity of the near-field criterion to the assumed effective radiating dimension.}
\label{tab:nearfield_sensitivity}
\setlength{\tabcolsep}{4pt}
\renewcommand{\arraystretch}{1.15}
\begin{tabularx}{\textwidth}{@{}
  >{\raggedright\arraybackslash}p{0.14\textwidth}
  >{\raggedright\arraybackslash}p{0.25\textwidth}
  >{\raggedright\arraybackslash}p{0.24\textwidth}
  >{\centering\arraybackslash}p{0.17\textwidth}
  >{\centering\arraybackslash}X
@{}}
\toprule
\textbf{Assumed \(D\)} &
\textbf{Physical basis} &
\textbf{Near-field criterion, \(r_{\mathrm{NF}}\)} &
\textbf{Rayleigh distance at \(105~\mathrm{kHz}\) (m)} &
\textbf{Implied \(f_{\mathrm{low}}\) (Hz)} \\
\midrule
\(0.20~\mathrm{m}\) &
Motor axial dimension &
\(\max(\lambda,\ 0.60,\ 2D^2f/c_w)\) &
\(5.5\) &
\({\sim}153\) \\

\addlinespace
\(0.40~\mathrm{m}\) &
Propeller diameter &
\(\max(\lambda,\ 1.20,\ 2D^2f/c_w)\) &
\(21.9\) &
\({\sim}153\) \\
\bottomrule
\end{tabularx}
\end{table}

For \(D_{\mathrm{eff}}=0.40~\mathrm{m}\) (propeller diameter), the Rayleigh distance \(r_{\mathrm{Ray}}=2D_{\mathrm{eff}}^2 f/c_w\) equals the \(10~\mathrm{m}\) CPA range at \(f_{\mathrm{Ray}}=r_{\mathrm{CPA}}c_w/(2D_{\mathrm{eff}}^2)=10\times1529/(2\times0.40^2)\approx47~\mathrm{kHz}\), and reaches \(r_{\mathrm{Ray}}=21.9~\mathrm{m}\) at the upper analysis limit of \(105~\mathrm{kHz}\). Source-related estimates at the highest analysis frequencies from the closest passages should therefore be interpreted with additional caution, because the far-field condition is only marginally satisfied for the propeller-diameter source extent at these frequencies. The wavelength-based transition frequency at the \(10~\mathrm{m}\) CPA geometry is \(f_{\lambda}=c_w/r_{\mathrm{CPA}}=1529/10\approx153~\mathrm{Hz}\). The adopted lower bound of \(200~\mathrm{Hz}\) therefore remains above the wavelength-based transition and is insensitive to the assumed compact-source dimension considered here. Consequently, the near-field assessment supports the selected lower-frequency bound, while the highest-frequency source-related estimates from the closest passages should be interpreted as propagation-corrected, source-related indicators rather than standardized far-field source levels.

\subsection{Shallow-Water Propagation Considerations}

In finite-depth water, low-frequency propagation is affected by normal-mode structure, seabed interaction, and waveguide cutoff effects~\cite{katsnelson2002shallow,jensen2011computational}. These effects can modify both received level and apparent spectral structure, introducing frequency-dependent bias into propagation-corrected estimates that is difficult to quantify without detailed environmental modelling. For the present survey area, with a mean water depth of approximately \(33~\mathrm{m}\), this consideration motivates caution at the low-frequency end of the analysis band and is a secondary motivation for the \(200~\mathrm{Hz}\) lower bound. Frequencies below \(200~\mathrm{Hz}\) may be examined qualitatively for diagnostic purposes but were not included in the principal source-related comparison.

\section{Doppler-Based Pass-Geometry Verification for the AUV test case}
\label{app:doppler}

Given an estimated source tonal frequency \(f_0\), AUV speed \(v\), 
sound speed \(c_w\), and a CPA range 
\(R_{\mathrm{CPA}}\), the Doppler-shifted frequencies at a stationary hydrophone can be predicted under a straight-line, constant-speed pass assumption~\cite{urick1983principles}. For the approaching and receding portions of the track,
\begin{equation}
f_{\mathrm{app}} = f_0 \frac{c_w}{c_w-v_r}, \qquad 
f_{\mathrm{rec}} = f_0 \frac{c_w}{c_w+v_r},
\label{eq:doppler_forward_appendix}
\end{equation}
where \(v_r\) is the radial component of the AUV speed relative to the receiver and \(c\) is the sound speed in water. If the tonal is evaluated symmetrically around the CPA at times \(\pm \Delta t\), then
\begin{equation}
x=v\Delta t, \qquad r=\sqrt{x^2+R_{\mathrm{CPA}}^2}, \qquad v_r = v\frac{x}{r} = v\frac{x}{\sqrt{x^2+R_{\mathrm{CPA}}^2}}.
\label{eq:vr_geom_compact}
\end{equation}
Because the pass geometry is symmetric around CPA under constant-speed straight-line motion, the Doppler shift before CPA must equal the Doppler shift after CPA, so that the frequency offsets from \(f_0\) are equal and opposite.

\begin{figure}[!htbp]
\centering
\includegraphics[width=0.9\textwidth]{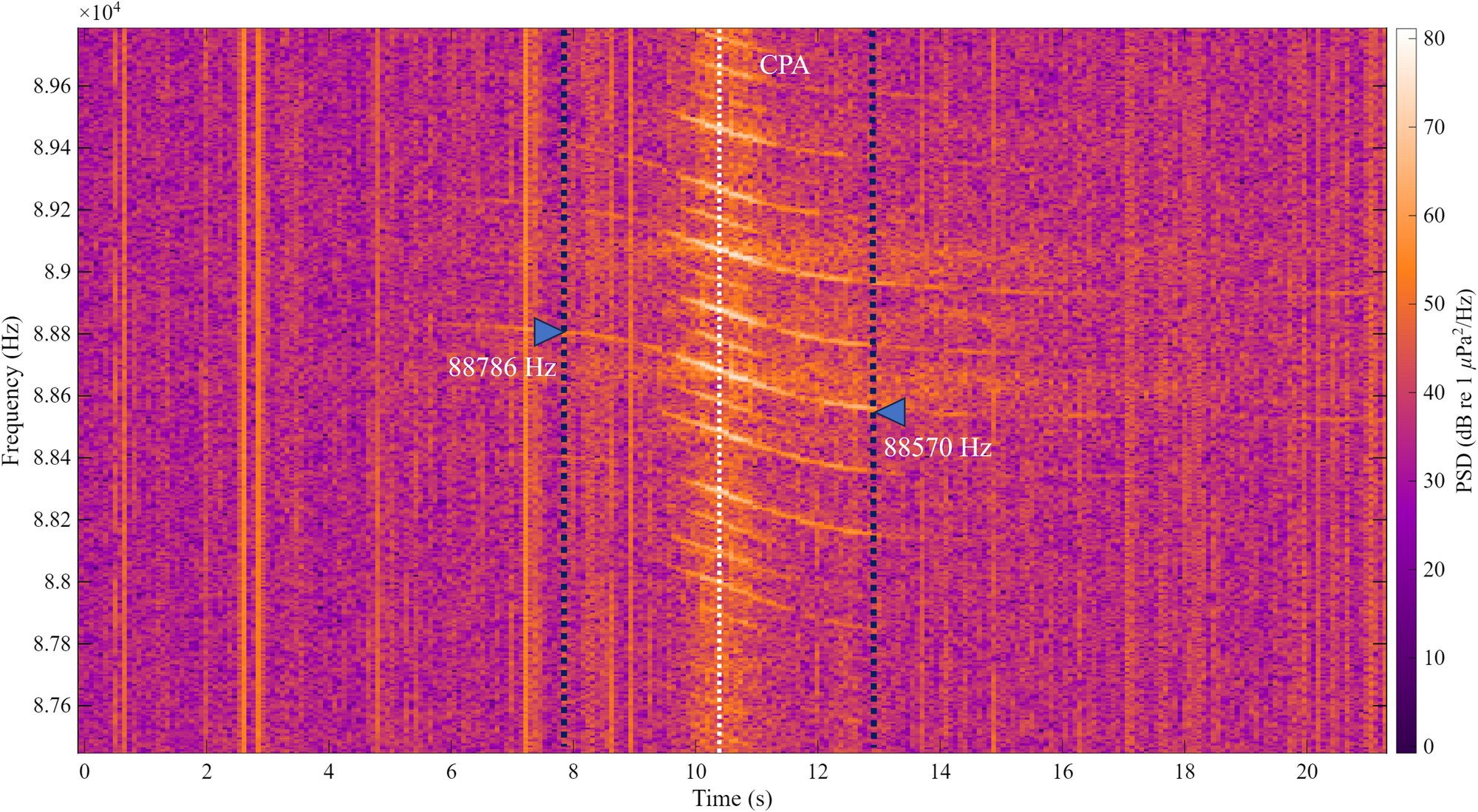}
\caption{Spectrum of the AUV around CPA (\(87{,}450\)--\(89{,}785~\mathrm{Hz}\), 100k-point FFT, Hann window, 50\% overlap). The white dashed line indicates CPA time, the dark blue dashed lines indicate the \(5~\mathrm{s}\) analysis window boundaries, and the observed frequencies are annotated on the plot.}
\label{fig:doppler}
\end{figure}

A spectrum of the AUV around the CPA is given in Fig.~\ref{fig:doppler}.

The Doppler analysis was performed using a \(5~\mathrm{s}\) window centered on CPA (\(\Delta t = 2.5~\mathrm{s}\)). The source tonal frequency was estimated as
\begin{equation}
f_0 = \frac{2f_{\mathrm{app,obs}}f_{\mathrm{rec,obs}}}{f_{\mathrm{app,obs}}+f_{\mathrm{rec,obs}}} \approx 88678~\mathrm{Hz}.
\end{equation}
The observed tonal frequencies were
\begin{equation}
f_{\mathrm{app,obs}} = 88786~\mathrm{Hz}, \qquad f_{\mathrm{rec,obs}} = 88570~\mathrm{Hz},
\end{equation}
corresponding to a symmetric shift of \(108~\mathrm{Hz}\) either side of \(f_0\). The implied radial speed is
\begin{equation}
v_r = c_w \cdot \frac{f_{\mathrm{app,obs}} - f_0}{f_{\mathrm{app,obs}}} \approx 1.862~\mathrm{m\,s^{-1}}.
\end{equation}

Using the geometric and environmental parameters \(v=2.8~\mathrm{m\,s^{-1}}\), \(R_{\mathrm{CPA}}=10~\mathrm{m}\), and \(c_w=1529.32~\mathrm{m\,s^{-1}}\), with \(\Delta t=2.5~\mathrm{s}\) giving \(x=7.0~\mathrm{m}\) and
\begin{equation}v_{r,\mathrm{pred}} = 2.8 \cdot \frac{7.0}{\sqrt{7.0^2+10.0^2}} \approx 1.606~\mathrm{m\,s^{-1}},
\end{equation}
The predicted frequencies at the nominal geometry are
\begin{equation}
f_{\mathrm{app,pred}} \approx 88771~\mathrm{Hz}, \qquad f_{\mathrm{rec,pred}} \approx 88585~\mathrm{Hz},
\end{equation}
corresponding to a predicted symmetric shift of \(93~\mathrm{Hz}\). The observed shift of \(108~\mathrm{Hz}\) exceeds this prediction by \(15~\mathrm{Hz}\), indicating a slightly larger Doppler spread than expected at the nominal pass geometry. Representative speed and CPA combinations consistent with the observed implied radial speed of \(1.862~\mathrm{m\,s^{-1}}\) are listed in Table~\ref{tbl:doppler_combinations}.

\begin{table}[!htbp]
\centering
\caption{Representative speed--CPA combinations consistent with the observed symmetric Doppler shift of \(108~\mathrm{Hz}\) in the \(5~\mathrm{s}\) window.}
\label{tbl:doppler_combinations}
\begin{tabular}{cc}
\toprule
Speed \(v\) (\(\mathrm{m\,s^{-1}}\)) & Equivalent \(R_{\mathrm{CPA}}\) (m) \\
\midrule
2.8 & 7.86 \\
2.9 & 8.72 \\
3.0 & 9.47 \\
3.1 & 10.32 \\
3.2 & 11.18 \\
3.5 & 13.93 \\
\bottomrule
\end{tabular}
\end{table}

The observed Doppler spread exceeding the nominal prediction is consistent with three possible explanations, which cannot be fully separated given the available data. First, the AUV speed during the pass may have been somewhat higher than the logged value of \(2.8~\mathrm{m\,s^{-1}}\), since DVL-derived speed estimates carry their own uncertainty and may not perfectly reflect the instantaneous vehicle speed relative to the ground. Although the AUV propulsion system supports speeds of up to approximately \(2.9~\mathrm{m\,s^{-1}}\) through the water, higher speeds over ground are achievable in the presence of favorable currents, and the speed combinations listed in Table~\ref{tbl:doppler_combinations} should be interpreted accordingly. Second, the true CPA range may have been shorter than the \(10~\mathrm{m}\) indicated by the AUV track data, since the reconstructed AUV trajectory is subject to DVL accuracy limitations, inertial navigation drift, and dead-reckoning error accumulation between surface GPS fixes. A shorter effective CPA range would increase the radial velocity component at a given time offset, leading to a larger observed Doppler spread. Third, limited precision in tracing the tonal ridge on the spectrogram may contribute to the residual, particularly at these high frequencies, where the tonal ridge may be narrow.

Because the CPA range is the primary geometric parameter used in the transmission-loss correction, the sensitivity of the source-related estimates to this uncertainty was evaluated explicitly in Sec.~\ref{subsubsec:source_related_tonal_psd}. That analysis showed that a \(\pm 1~\mathrm{m}\) variation in CPA range produces TL deviations of up to approximately \(\pm 1.5~\mathrm{dB}\), which is included in the stated combined geometric uncertainty. The nominal logged CPA range of \(\sim 10~\mathrm{m}\) is retained for the source-related estimates, and the residual Doppler deviation is noted as a contributing factor to the overall geometric uncertainty rather than as evidence of a systematic range bias.

\section{Image-Source Propagation Model Configuration}
\label{app:image_source}

Transmission-loss corrections for the source-related tonal PSD estimates in 
Section~\ref{subsubsec:source_related_tonal_psd} were computed using a 
close-range image-source propagation model accounting for direct, 
surface-reflected, and bottom-reflected paths~\cite{ainslie2022image}. The 
model was configured to match the environmental parameters observed on the day 
of the survey, using a sound speed of $c = 1529~\mathrm{m\,s^{-1}}$, a water 
depth of $H = 33~\mathrm{m}$, a source depth of $z_s = 20~\mathrm{m}$, and a 
maximum reflection order of~4. The sea surface was modeled with a reflection 
coefficient magnitude $|R_s| = 0.9$ and a phase of $180^{\circ}$, 
approximating a near-pressure-release boundary consistent with the calm sea 
state ($H_s = 0$--$0.2~\mathrm{m}$) observed during the survey. The seabed 
was modeled with a reflection coefficient magnitude $|R_b| = 0.7$ and a phase 
of $0^{\circ}$, representative of a sandy-silt sediment at the survey site. TL 
was computed in band-averaged intensity mode over $N_f = 25$ linearly spaced 
frequency samples per band, on a range grid of $1$--$100~\mathrm{m}$ 
($N_r = 500$) and a depth grid of $N_z = 250$ points. Frequency-dependent 
absorption was evaluated per band using the Thorp 
formula~\cite{thorp1967analytic}, yielding values in the range 
$0.001$--$15.2~\mathrm{dB\,km^{-1}}$ across the analysis band.

Consistent with~\eqref{eq:tl_bandwidth}, TL was energetically averaged over 
finite frequency bands rather than evaluated at individual FFT bins, to reduce 
sensitivity to local interference artifacts. The averaging bands were 
$200$--$1000~\mathrm{Hz}$, $1000$--$2000~\mathrm{Hz}$, 
$2000$--$4000~\mathrm{Hz}$, $4000$--$8000~\mathrm{Hz}$, followed by 
successive $4000~\mathrm{Hz}$-wide bands up to the upper analysis limit. A 
finer band resolution was used at lower frequencies, where shallow-water 
propagation and seabed interaction can produce stronger frequency-dependent 
variability; all bands are wider than the modal-spacing reference 
of~\eqref{eq:tl_bandwidth}, and the TL corrections therefore represent 
band-averaged modal energies~\cite{katsnelson2002shallow,jensen2011computational}.

\begin{figure}[!htbp]
\centering
\includegraphics[width=0.8\textwidth]{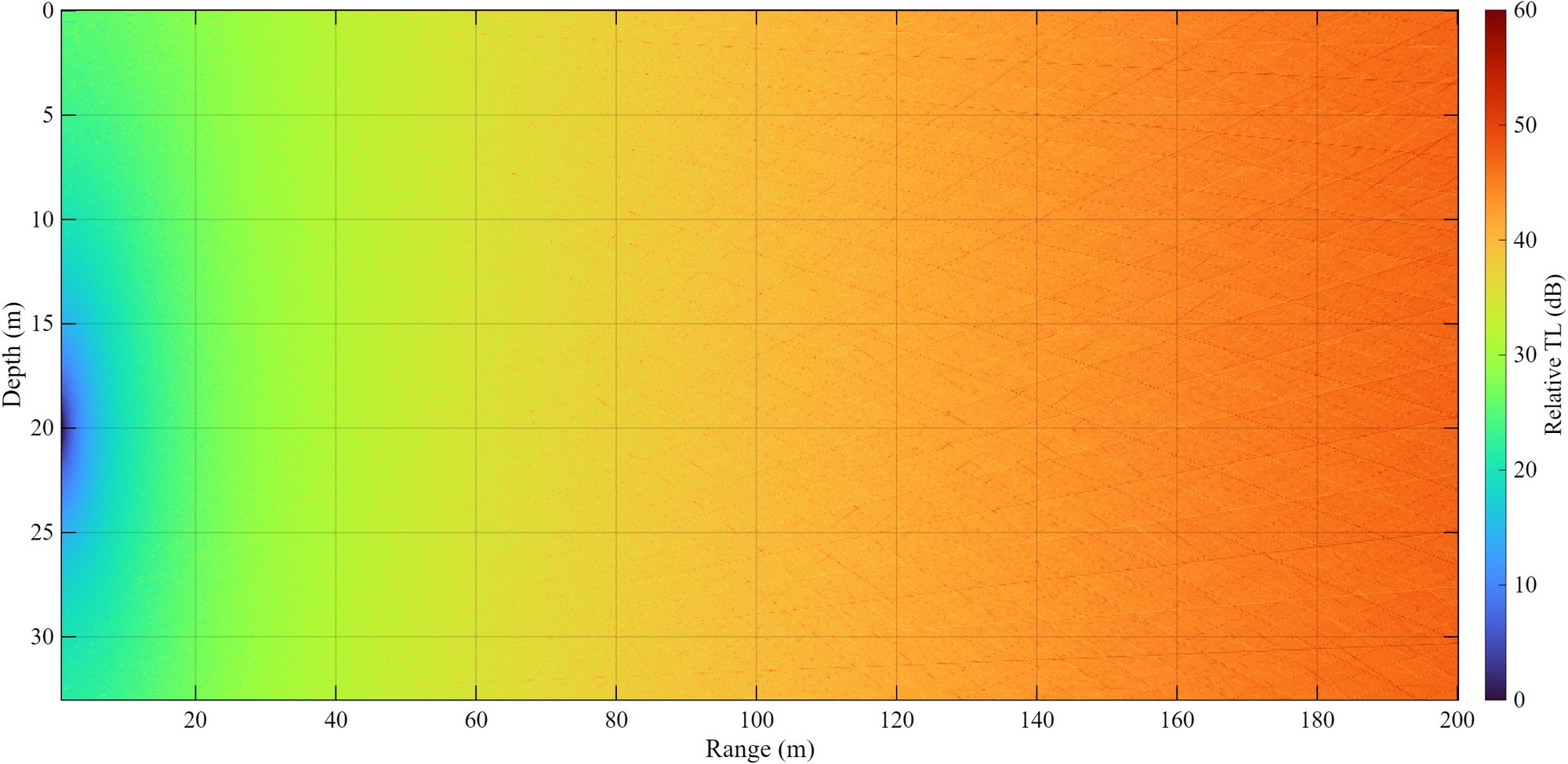}
\caption{Transmission-loss field computed using the image-source propagation 
model. Results are shown over the $200~\mathrm{Hz}$--$105~\mathrm{kHz}$ 
analysis band and a range of $1$--$200~\mathrm{m}$, at a source depth of 
$z_s = 20~\mathrm{m}$, water depth $H = 33~\mathrm{m}$, sound speed 
$c = 1529~\mathrm{m\,s^{-1}}$, surface reflection coefficient $|R_s| = 0.9$ 
($180^{\circ}$), and seabed reflection coefficient $|R_b| = 0.7$ ($0^{\circ}$).
Band-averaged TL corrections were evaluated separately for each analysis band 
as described in the text.}
\label{fig:tl_profile}
\end{figure}

The resulting TL field is shown in Fig.~\ref{fig:tl_profile}.

\clearpage

\bibliographystyle{elsarticle-num}
\bibliography{references}

\end{document}